\documentclass[fleqn,10pt]{wlscirep}
\usepackage[utf8]{inputenc}
\usepackage[T1]{fontenc}
\usepackage{longtable}
\usepackage[resetlabels]{multibib}
\newcites{Supp}{References}

\usepackage[normalem]{ulem}
\newcommand{\nzradd}[1]{#1}
\newcommand{\nzrremove}[1]{}

\title{A close-in giant planet escapes engulfment by its star}

\author[*1,35]{Marc Hon}
\author[1,2]{Daniel Huber}
\author[3]{Nicholas Z. Rui}
\author[3]{Jim Fuller}
\author[4,5,6]{Dimitri Veras} 

\author[7,8]{James S. Kuszlewicz}
\author[9]{Oleg Kochukhov}
\author[8, 10, 11]{Amalie Stokholm}
\author[8]{Jakob Lysgaard R\o{}rsted}
\author[12]{Mutlu Yıldız}
\author[12]{Zeynep \c{C}elik Orhan}
\author[12]{Sibel \"{O}rtel}
\author[13]{Chen Jiang} 

\author[1]{Daniel R. Hey}
\author[14]{Howard Isaacson}
\author[1]{Jingwen Zhang}
\author[15]{Mathieu Vrard}
\author[16]{Keivan G. Stassun}
\author[1]{Benjamin J. Shappee}
\author[17,1]{Jamie Tayar}
\author[17,1]{Zachary R. Claytor} 

\author[18]{Corey Beard}
\author[2]{Timothy R. Bedding}
\author[1]{Casey Brinkman}
\author[19,20]{Tiago L. Campante}
\author[21]{William J. Chaplin}
\author[22,1]{Ashley Chontos}
\author[13]{Steven Giacalone}
\author[17]{Rae Holcomb}
\author[23]{Andrew W. Howard}
\author[17]{Jack Lubin}
\author[24]{Mason MacDougall}
\author[25, 26, 27]{Benjamin T. Montet}
\author[28]{Joseph M. A. Murphy}
\author[1,32,35]{Joel Ong}
\author[29]{Daria Pidhorodetska}
\author[30]{Alex S. Polanski}
\author[31, 32]{Malena Rice}
\author[25,2,33]{Dennis Stello}
\author[24]{Dakotah Tyler}
\author[24]{Judah Van Zandt}
\author[34]{Lauren Weiss} 

\affil[1]{Institute for Astronomy, University of Hawai`i,
2680 Woodlawn Drive, Honolulu, HI 96822, USA}
\affil[2]{Sydney Institute for Astronomy (SIfA), School of Physics, University of Sydney, Camperdown, NSW 2006, Australia}

\affil[3]{TAPIR, Mailcode 350-17, California Institute of Technology, Pasadena, CA 91125, USA}
\affil[4]{Centre for Exoplanets and Habitability, University of Warwick, Coventry CV4 7AL, UK}
\affil[5]{Centre for Space Domain Awareness, University of Warwick, Coventry CV4 7AL, UK}
\affil[6]{Department of Physics, University of Warwick, Coventry CV4 7AL, UK}
\affil[7]{Center for Astronomy (Landessternwarte), Heidelberg University, K\"onigstuhl 12, 69118 Heidelberg, Germany}
\affil[8]{Stellar Astrophysics Centre, Department of Physics and Astronomy, Aarhus University, Ny Munkegade 120, DK-8000 Aarhus C, Denmark}
\affil[9]{Department of Physics and Astronomy, Uppsala University, Box 516, Uppsala 75120, Sweden}
\affil[10]{Dipartimento di Fisica e Astronomia, Universit
\`{a} degli Studi di Bologna, Via Gobetti 93/2, I-40129 Bologna, Italy}
\affil[11]{INAF – Osservatorio di Astrofisica e Scienza dello Spazio di Bologna, Via Gobetti 93/3, I-40129 Bologna, Italy}
\affil[12]{Department of Astronomy and Space Sciences, Science Faculty, Ege University, 35100, Bornova,  $\dot{\mathrm{I}}$zmir, Turkey}
\affil[13]{Max-Planck-Institut f{\"u}r Sonnensystemforschung, Justus-von-Liebig-Weg 3, 37077 G{\"o}ttingen, Germany}
\affil[14]{Department of Astronomy, University of California Berkeley, Berkeley CA 94720, USA}
\affil[15]{Department of Astronomy, The Ohio State University, 140 West 18th Avenue, Columbus OH 43210, USA}
\affil[16]{Department of Physics and Astronomy, Vanderbilt University, Nashville, TN 37235, USA}
\affil[17]{Department of Astronomy, University of Florida, Bryant Space Science Center, Stadium Road, Gainesville, FL 32611, USA}

\affil[18]{Department of Physics and Astronomy, University of California Irvine, Irvine, CA 92697, USA}

\affil[19]{Instituto de Astrof\'{i}sica e Ci\^{e}ncias do Espa\c{c}o, Universidade do Porto, CAUP, Rua das Estrelas, 4150-762 Porto, Portugal}
\affil[20]{Departamento de F\'{i}sica e Astronomia, Faculdade de Ci\^{e}ncias da Universidade do Porto, Rua do Campo Alegre, s/n, 4169-007
Porto, Portugal}
\affil[21]{School of Physics and Astronomy, University of Birmingham, Edgbaston, Birmingham B15 2TT, United Kingdom}
\affil[22]{Department of Astrophysical Sciences, Princeton University, 4 Ivy Lane, Princeton, NJ 08544, USA}
\affil[23]{Department of Astronomy, California Institute of Technology, Pasadena, CA 91125, USA}
\affil[24]{Department of Physics and Astronomy, University of California Los Angeles, Los Angeles, CA 90095, USA}
\affil[25]{School of Physics, University of New South Wales, Sydney, NSW 2052, Australia}
\affil[26]{UNSW Data Science Hub, University of New South Wales, Sydney, NSW, 2052, Australia}
\affil[27]{Australian Centre for Astrobiology, University of New South Wales, Sydney, NSW, 2052, Australia}

\affil[28]{Department of Astronomy and Astrophysics, University of California, Santa Cruz, CA 95064, USA}
\affil[29]{Department of Earth and Planetary Sciences, University of California, Riverside, CA 92521, USA}
\affil[30]{Department of Physics and Astronomy, University of Kansas, 1082 Malott, 1251 Wescoe Hall Dr., Lawrence, KS 66045, USA}
\affil[31]{Department of Physics and Kavli Institute for Astrophysics and Space Research, Massachusetts Institute of Technology, Cambridge, MA 02139, USA}
\affil[32]{Department of Astronomy, Yale University, New Haven, CT 06511, USA}

\affil[33]{ARC Centre of Excellence for All Sky Astrophysics in Three Dimensions (ASTRO-3D)}

\affil[34]{Department of Physics and Astronomy, University of Notre Dame, Notre Dame, IN 46556}
\affil[35]{NASA Hubble Fellow}

\begin{abstract}
 When main-sequence stars expand into red giants, they are expected to engulf close-in planets \cite{Nordhaus_2013, Madappatt_2016,Gallet_2017, Ronco_2020, Grunblatt_2022}. Until now, the absence of planets with short orbital periods around post-expansion, core helium-burning red giants \cite{Sato_2008, Kunitomo_2011, Villaver_2014} has been interpreted as evidence that short-period planets 
around Sun-like stars do not survive the giant expansion phase of their host stars \cite{Macleod_2018}.
Here, we present the discovery that the giant planet 8 Ursae Minoris b \cite{Lee_2015} orbits a core helium-burning red giant. At a distance of only 0.5 au from its host star, the planet would have been engulfed by its host star, which is predicted by standard single-star evolution to have previously expanded to a radius of 0.7 au. Given the brief lifetime of helium burning giants, 
the planet’s nearly circular orbit is challenging to reconcile with scenarios whereby the planet survives by having a distant orbit initially.  Instead, the planet may have avoided engulfment through the scenario of a stellar merger that either altered the evolution of the host star or produced 8 Ursae Minoris b as a second generation planet \cite{Perets_2011}. This system shows that core helium-burning red giants can harbour close planets and provides evidence for the role of non-canonical stellar evolution in the extended survival of late-stage exoplanetary systems.

\end{abstract}
\begin{document}

 

\flushbottom
\maketitle

\begin{figure}
\centering
\includegraphics[width=\linewidth]{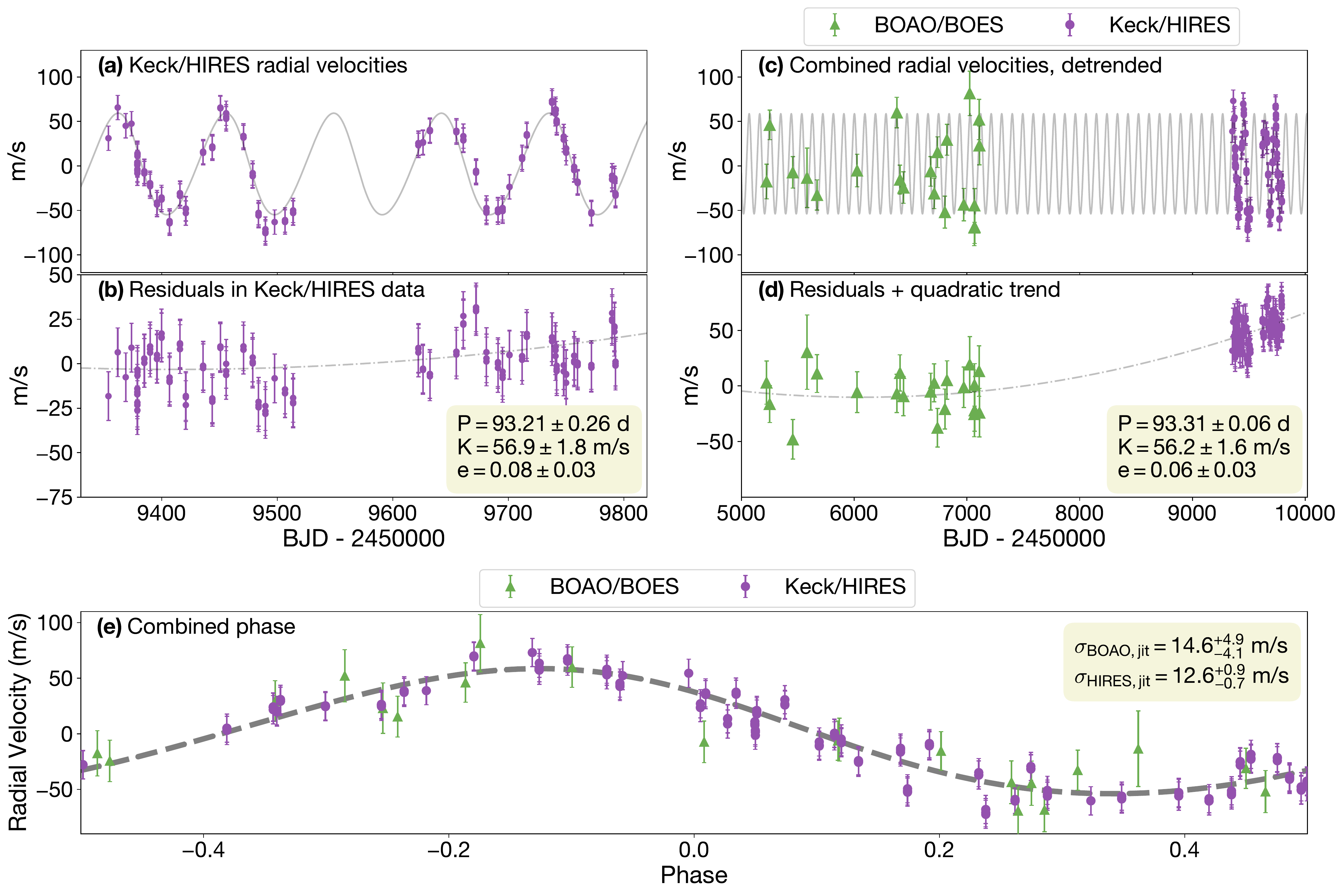}
\caption{\textbf{Radial velocity measurements of host star 8 UMi}. (\textit{a}) Radial velocity time series from Keck/HIRES observations. The
curve shows the best fit to the radial velocity measurements whose error bars
are 1$\sigma$ measurement uncertainties added in quadrature with a fitted jitter
value.  (\textit{b}) Residuals after subtracting the best fit from the time series, fitted with a quadratic trend in time as shown by the dash-dotted line. (\textit{c}) 
Combined observations from Keck/HIRES and BOAO/BOES. A quadratic trend in time has already been fitted and subtracted to the data shown here to show variations from planet 8 UMi b alone. (\textit{d}) Residuals in the radial velocity data after subtracting the best fit indicated by the curve in panel (\textit{c}).
(\textit{e}) The phase-folded time series of the combined and detrended radial velocity data. The dashed line shows the phase-folded time series of the best fit to the orbit presented in panel (\textit{c}).  }
\label{fig:2}
\end{figure}

The red giant 8 Ursae Minoris (hereafter 8 UMi), known also as Baekdu, was discovered to host the giant planet 8 UMi b (known also as Halla) on a close-in, near-zero eccentricity ($e\simeq0.06$) orbit with a period of $93.4\pm4.5$ days
based on observations from the Bohyunsan Optical Astronomy Observatory (BOAO) \cite{Lee_2015}. Radial velocity detections of close-in planets surrounding red giants are sometimes ambiguous \cite{Hatzes_2018, Dollinger_2021}, and so we confirmed 8 UMi b's detection by collecting 135 additional radial velocity measurements using the HIRES spectrograph \cite{Vogt_1994} on the Keck-I telescope at Maunakea, Hawaii. A Keplerian orbit fit to the combined radial velocity data refines the planet's estimated orbital properties to a period of $93.31\pm0.06$ days and an eccentricity of $0.06\pm0.03$ (Table \ref{table:mcmc}), demonstrating phase coherence of the radial velocity data across 12.5 years or 49 orbital cycles (Fig. \ref{fig:2}). A trend visible within the residuals of the combined fit suggests an additional, outer companion in the planetary system, whose minimum distance was determined with a 67\% confidence to be 5 au.

The long term stability of 8 UMi's 93-day radial velocity variations suggests an orbiting companion rather than stellar activity modulated by the star's rotation. In support of this, we have found no evidence of a 93-day photometric variability for the host star. Additional tests of chromospheric and surface magnetic fields do not detect significant activity levels expected for a red giant rotating with a period consistent with the RV data. These tests, which are detailed in \textbf{Methods}, support the planetary nature of the radial velocity variations.

\begin{figure}
\centering
\includegraphics[width=\linewidth]{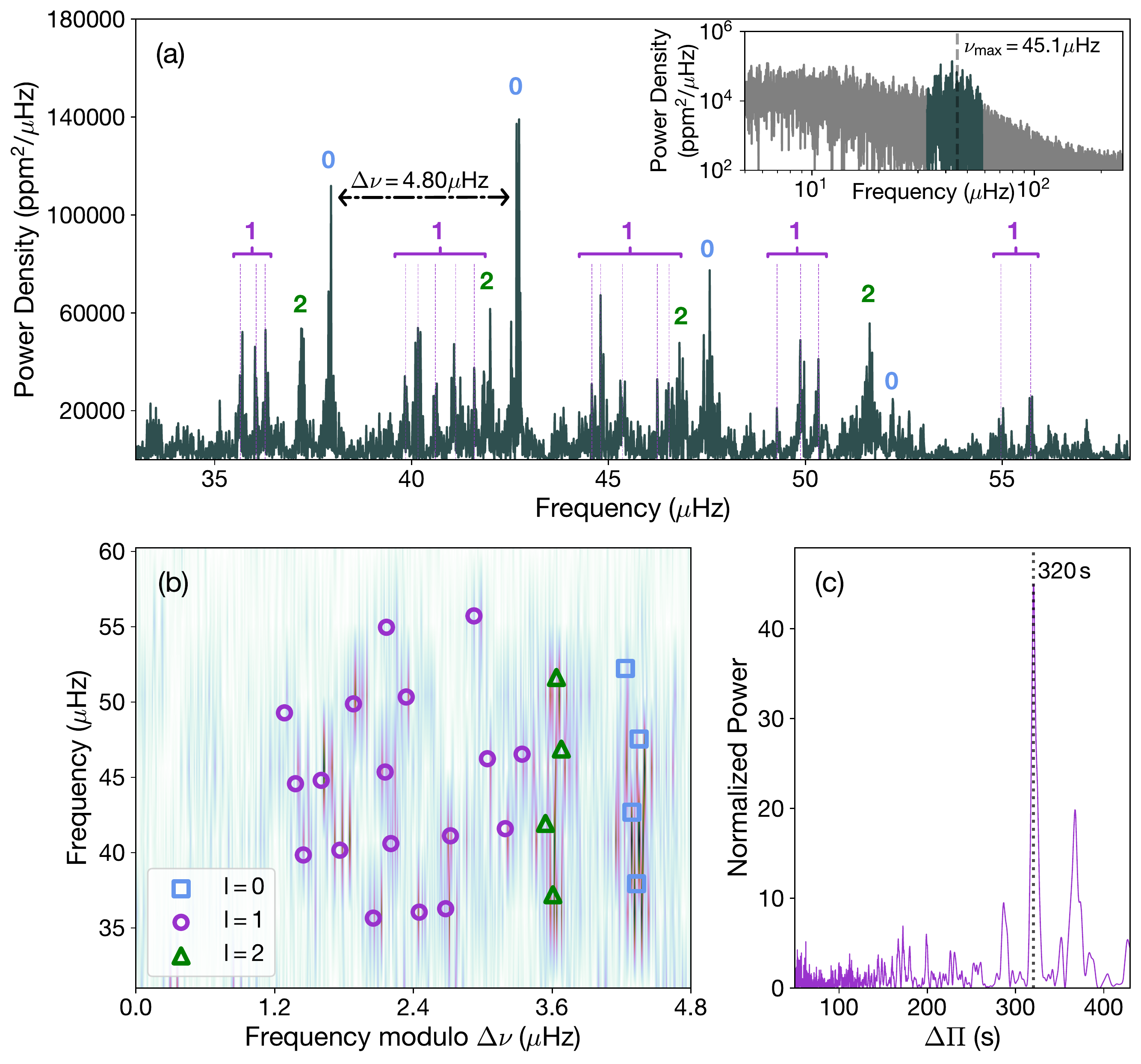}
\caption{\textbf{Identification of the evolutionary state of host star 8 UMi as a core helium-burning red giant}. (\textit{a}) Oscillation power spectrum of 8 UMi, as observed by TESS. Oscillation modes are labelled by their angular degree $l$. The inset presents a zoomed out view of the spectrum. (\textit{b}) The frequency \'echelle diagram of the spectrum. The symbols indicate the labelled oscillation modes. (\textit{c}) Fourier transform of the stretched period spectrum for the $l=1$ modes (see \textbf{Methods}). The 
asymptotic $l=1$ period spacing value for 8 UMi corresponds to the value of $\Delta\Pi$ with the largest power, which is 320$\,$s.}
\label{fig:1}
\end{figure}

8 UMi was observed by NASA's Transiting Survey Satellite (TESS) for 6 months between July 2019 and June 2020, and for another 6 months between June 2021 and June 2022. The star's oscillation modes are well-resolved within the TESS data (Fig. \ref{fig:1}a-b), such that its evolutionary state can be determined using asteroseismology. The period spacing $\Delta\Pi$ of the seismic dipole ($l=1$) oscillation modes for red giants can distinguish between red giants that are expanding during their first ascent in the Hertzsprung-Russell diagram and those that have previously swelled to their maximum size, but now reside in the so-called red clump with smaller sizes \cite{Bedding_2011}. First ascent red giants burn hydrogen in a shell surrounding an inert helium core, resulting in a strong density gradient between the contracting core and envelope, with $\Delta\Pi$ in the range $50-100$\,s. Meanwhile, red clump stars burn helium within an expanded, convective core, and have $\Delta\Pi$ in the range $250-400$\,s \cite{Vrard_2016}. We measured $\Delta\Pi = 320$\,s for 8 UMi, which unambiguously identifies the star as a core-helium burning giant (Fig. \ref{fig:1}c).

To determine the mass of the host star, we used established stellar modelling techniques that compare the star's observed properties to predicted observables from stellar evolution models. Using observables from spectroscopy and asteroseismology, we estimated the star's mass to be $1.51 \pm 0.05\,M_{\odot}$, which translates into a 
semi-major axis of $0.462 \pm 0.006\,$au and a minimum planetary mass of $1.65\pm0.06\,M_J$ for 8 UMi b.
Based on the host star's mass and metallicity, stellar models predict that 8 UMi would have once expanded to about 0.7 AU. Depending on the choice of model, we confirm with a 3--8$\sigma$ confidence that the expanding host star would have surpassed the planet's current orbital distance (Fig. \ref{fig:3}).

While brown dwarfs \cite{Maxted_2006} or low-mass stars may survive host star engulfment in common-envelope scenarios, 8 UMi b is less massive than a brown dwarf with a 99.2\% probability, assuming orbits are oriented randomly. Additionally, astrometric data from ESA's \textit{Gaia} mission precludes unresolved companions more massive than red dwarfs in a face-on orbit. Irrespective of the companion's mass, a common-envelope event with the host star is unlikely because the resulting gravitational potential energy release could not have ejected the host star's envelope (see \textbf{Methods}). Additionally, the engulfed companion would need only $10^2-10^4$ yr to spiral \cite{Macleod_2018} into the star, in contrast to the Myr timescale required for the star to complete its red giant expansion.

One possibility explaining 8 UMi b's existence is that the planet initially orbited at larger distances, but was pulled inwards by tidal interactions from its expanding host star. However, simulations \cite{Villaver_2009, Kunitomo_2011, Villaver_2014} indicate that this scenario would require an exceptionally fine-tuned migration to halt further orbital decay into the stellar envelope and reproduce the planet's near-zero eccentricity orbit.
Alternatively, dynamical interactions may have scattered the planet inwards on an eccentric orbit \cite{Rasio_1996}, which then tidally circularized. However, the planet is then required to have attained its current, near-circular orbit only after the host star has reached its maximum size and commenced helium burning.  
Tidal circularization of the planet's orbit would thus need to occur within the 100 Myr that the host star spends in its core helium-burning stage, which is unlikely given that
the planet's orbit requires at minimum several Gyr to circularize at its current distance from its host star.

\begin{figure}
\centering
\includegraphics[width=0.6\linewidth]{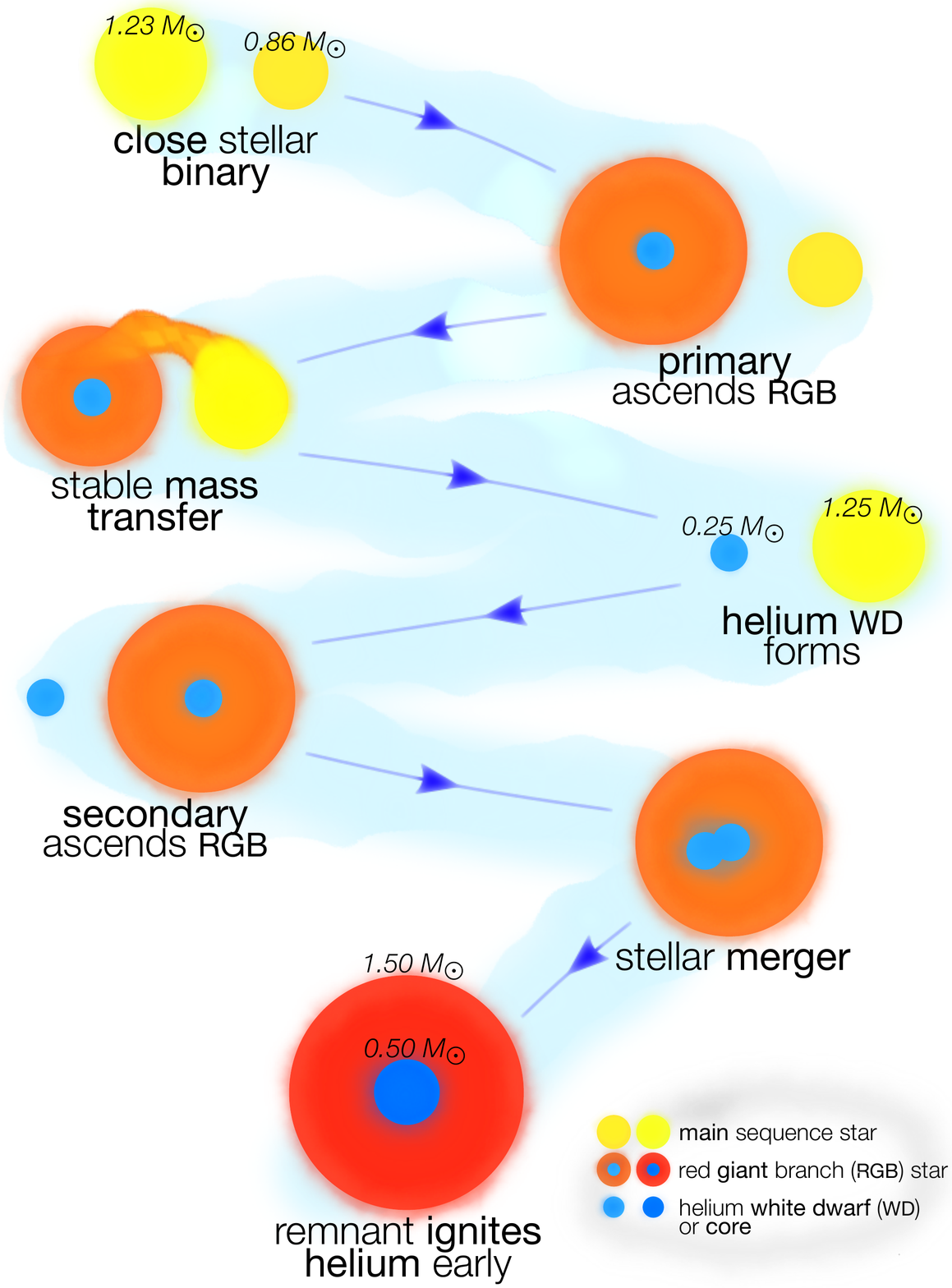}
\caption{\textbf{A possible history of host star 8 UMi in the form of a stellar merger}.
A primary, upon its ascent up the red giant branch (RGB), is first stripped by its companion into a helium white dwarf (WD). The white dwarf merges with the secondary when it evolves off the main sequence and ascends the RGB.
This binary scenario produces a $M\approx1.5M_\odot$ core helium burning star without requiring the star's radius to grow beyond the orbital distance of 8 UMi b.}
\label{fig:luckyplanet_cartoon}
\end{figure}

Models of binary stellar evolution \cite{Hurley_2002, Izzard_2007} show that white dwarf--red giant mergers can ignite core helium, resulting in the early termination of a red giant's expansion up the giant branch. Using Modules for Experiments in Stellar Astrophysics \cite{paxton2010modules_main}, we modelled a binary history for 8 UMi and verified that a circumbinary planet can remain dynamically stable throughout the binary evolution of two low-mass stars that initially orbit each other at a 2-day period before merging once forming a white dwarf--red giant pair.
Therefore, 8 UMi may have been a close binary system whose merger prevented its components from expanding sufficiently to engulf its circumbinary planet (Fig. \ref{fig:luckyplanet_cartoon}). After the merger, 8 UMi is expected to evolve as a typical red clump star, such that the planet will eventually be engulfed once the star exhausts core helium and expands up the asymptotic giant branch. According to this scenario, 8 UMi's progenitor once was an EL CVn-type binary \cite{Maxted_2014}. These are binary systems with an A- or F-type star orbiting a very low-mass white dwarf at periods shorter than 3 days. 
Such systems frequently form the inner binaries of triple star systems \cite{Tokovinin_2006, Lagos_2020}, suggesting a stellar nature for the unresolved outer companion in the 8 UMi system. A binary evolution scenario for 8 UMi is further supported by predictions of surface lithium enrichment for the remnant core-helium burning star based on simulations \cite{zhang2013white, Zhang_2020} of white dwarf--red giant mergers. In particular, measurements from previous spectroscopic surveys \cite{Kumar_2011, Charbonnel_2020} have indicated that host star 8 UMi is over-abundant in lithium, with $\mathrm{A(Li)}= 2.0\pm0.2$. With a 3$\sigma$ confidence, this is a factor of ten to a hundred times larger than the lithium abundances of observed red clump stars and single star model predictions at this mass \cite{Magrini_2021,Chaname_2022}.
Besides a circumbinary planet history, another possibility is that 8 UMi b formed as a byproduct of the stellar merger event. Stellar ejecta from binary interactions have been hypothesized to form protoplanetary disks \cite{Perets_2011}, and so 8 UMi b may have emerged from the debris disk surrounding the merger remnant.

As the first known close-in planet around a helium-burning red
giant, 8 UMi b demonstrates that stellar multiplicity, whose effect influences the fates of planetary systems \cite{Kraus_2016, Moe_2021}, may result in pathways through which planets survive the volatile evolution of their parent stars. Given the ubiquity of binary systems, there may be a greater preponderance of exoplanets orbiting post-main sequence host stars than previously assumed. 





\begin{table}[ht]
\centering
\caption{\textbf{Parameters of the fitted Keplerian orbit to the 8 UMi system}. The median values of the posterior distribution are reported. The lower- and upper-bound
uncertainties are the intervals between the median with the 16$^{\mathrm{th}}$ and 84$^{\mathrm{th}}$ percentile values of the distribution, respectively. BOES, Bohyunsan Observatory Echelle Spectrograph.;
BJD, barycentric Julian date.}\label{table:mcmc}
\begin{tabular}{lr}
\toprule
\multicolumn{2}{c}{\textbf{Orbital Parameters}} \\
\midrule
Orbital period, $P$ & $93.31 \pm 0.06$ days \\
Radial velocity semi-amplitude, $K$ & $56.1^{+1.7}_{-1.6}$ m/s \\
Eccentricity, $e$ & $0.062^{+0.028}_{-0.030}$\\
Argument of periapsis, $\omega$ & $0.942^{+0.533}_{-0.463}$ rad\\
Time of periastron passage (BJD), $t_P$ & $2457601.379^{+7.844}_{-6.936}$\\
\midrule
\multicolumn{2}{c}{\textbf{Instrumental Parameters}} \\ 
\midrule
BOAO/BOES center of mass velocity, $\gamma_{\mathrm{BOES}}$ & $13.33^{+7.42}_{-7.49}$ m/s\\
Keck/HIRES center of mass velocity, $\gamma_{\mathrm{HIRES}}$ & $-46.91 \pm 11.92$ m/s\\
BOAO/BOES jitter, $\sigma_{\mathrm{BOES}}$ & $14.59^{+4.89}_{-4.12}$ m/s\\
Keck/HIRES jitter, $\sigma_{\mathrm{HIRES}}$ & $12.65^{+0.86}_{-0.75}$ m/s\\
Linear acceleration, $\dot{\gamma}$ & $0.014\pm0.005$ m/s d$^{-1}$\\
Curvature ($\times10^{-5}$), $\ddot{\gamma}$ & $0.49\pm0.15$ m/s d$^{-2}$\\
\midrule
\multicolumn{2}{c}{\textbf{Derived Parameters}} \\
\midrule
Planet mass, $M_p\sin i$ & $1.65\pm0.06$ $M_J$\\
Planet semi-major axis, $a_{\mathrm{pl}}$ & $0.462\pm0.006$ au\\
\bottomrule
\end{tabular}
\end{table}


\clearpage

\section*{Acknowledgements}
The authors wish to recognize and acknowledge the very significant cultural role and reverence that the summit of Maunakea has within the indigenous Hawaiian community.  We are most fortunate to have the opportunity to conduct observations from this mountain. The data in this study were obtained at the W. M. Keck Observatory, which is operated as a scientific partnership among the California Institute of Technology, the University of California and the National Aeronautics and Space Administration. The Observatory was made possible by the generous financial support of the W. M. Keck Foundation. Additional observations were obtained at the Canada-France-Hawaii Telescope (CFHT), which is operated by the National Research Council (NRC) of Canada, the Institut National des Sciences de l'Univers of the Centre National de la Recherche Scientifique (CNRS) of France, and the University of Hawaii.
M.H. acknowledges support from NASA through the NASA Hubble
Fellowship grant HST-HF2-51459.001 awarded by the Space Telescope
Science Institute, which is operated by the Association of Universities for Research in Astronomy, Incorporated, under NASA contract NAS5-26555.
D.H. acknowledges support from the Alfred P. Sloan Foundation, the National Aeronautics and Space Administration (80NSSC21K0652, 80NSSC20K0593), and the Australian Research Council (FT200100871).
N.Z.R. acknowledges support from the National Science Foundation Graduate Research Fellowship under Grant No. DGE‐1745301.
A.S. acknowledges support from the European Research Council Consolidator Grant funding scheme (project ASTEROCHRONOMETRY, G.A. n. 772293).
O.K. acknowledges support from the Swedish Research Council under the project grant 2019-03548.
M.V. acknowledges support from NASA grant 80NSSC18K1582.
This work was supported by Funda\c c\~ao para a Ci\^encia e a Tecnologia (FCT) through research grants UIDB/04434/2020 and UIDP/04434/2020. T.L.C.~is supported by FCT in the form of a work contract (CEECIND/00476/2018).
T.R.B. acknowledges support from the Australian Research Council through Discovery Project DP210103119 and Laureate Fellowship FL220100117.

\section*{Author contributions statement}
M.H. identified 8 UMi's oscillations, led the observing program and data analysis, and wrote most of the manuscript.
D.H. organized observations, interpreted the asteroseismic and radial velocity data, and contributed to writing the manuscript.
N.Z.R. and J.F. conducted binary simulations for the host star, performed numerical calculations for planet survival scenarios, and contributed to writing the manuscript.
J.F. and D.V. interpreted formation scenarios for the host star.
J.S.K. and M.V. extracted oscillation parameters from the TESS data. 
O.K. performed the spectropolarimetric analysis of the host star and the control target. A.S., J.L.R., M.Y., Z.\c{C}.O., S.\"{O}., C.J., and J.O. conducted grid-based modelling for 8 UMi.
D.R.H. D.H., and M.H. fitted the radial velocity data.
H.I. measured chromospheric activity indices from the HIRES data.
J.Z. constrained the properties of the outer companion.
K.G.S. performed the SED analysis for the host star.
B.J.S. extracted ASAS-SN photometry for 8 UMi.
J.T. and Z.R.C. provided interpolatable grids of isochrones.
T.R.B. and D.S. analyzed the asteroseismic data and helped guide the strategy of the manuscript. 
B.T.M. identified and analyzed the lithium richness of the control target.
W.J.C., D.H., and T.L.C., are key architects of TASC working groups on exoplanet hosts, including evolved stars.  
H.I. and A.W.H. oversaw the California Planet Search observing program.
A.C., S.G., C.B., J.L., R.H., J.M.A.M., J.V.Z., D.T., D.P., C.B., M.M. A.S.P., M.R., and L.W. conducted Keck I/HIRES observations of 8 UMi and the control star.
All authors reviewed the manuscript.

\section*{Competing interests}
The authors declare that they have no competing financial interests.

\section*{Correspondence}
Correspondence and requests for materials should be addressed to M.H. (email: mtyhon@hawaii.edu).

\begin{figure}
\centering
\includegraphics[width=\linewidth]{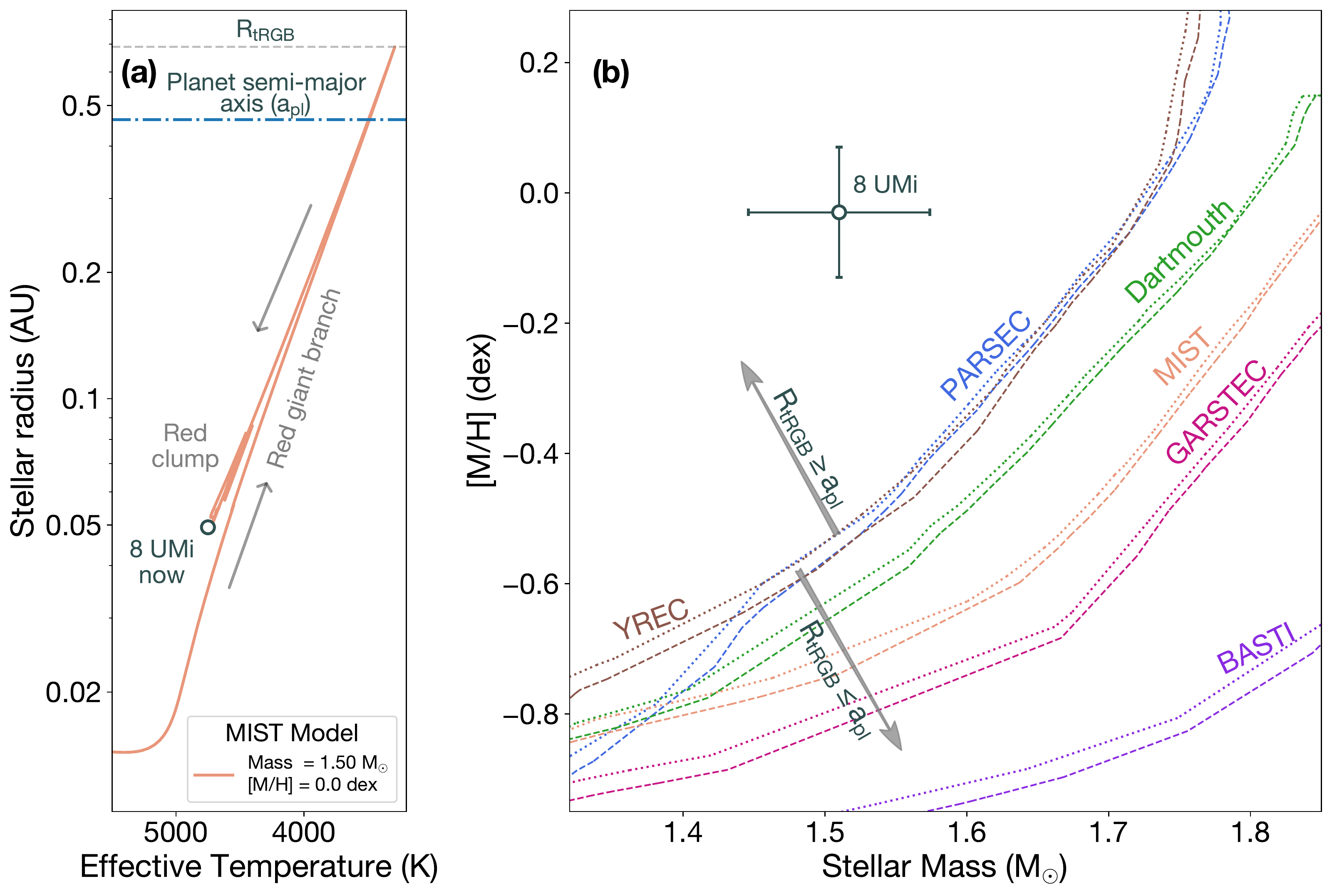}
\caption{\textbf{Canonical single-star evolutionary models for the host star 8 UMi.} (\textit{a}) A 1.5$M_{\odot}$, solar-metallicity stellar model indicates that host star 8 UMi would have reached a maximum size of $R_{\mathrm{tRGB}}$ during its first ascent up the red giant branch. This maximum size exceeds the semi-major axis of the planet 8 UMi b, $a_{\mathrm{pl}} = 0.462 \pm 0.006$ au (dashed-dotted line). (\textit{b}) Boundaries at which $R_{\mathrm{tRGB}} = a_{\mathrm{pl}}$ are shown across a range of stellar mass and metallicity ([M/H]) values for different stellar evolutionary codes. Each boundary comprises two lines, which reflect the $\pm1\sigma$ uncertainties of $a_{\mathrm{pl}}$. The mass and metallicity measurements for 8 UMi, whose error bars indicate 1$\sigma$ (standard deviation) uncertainties, locate the star in the region where $R_{\mathrm{tRGB}} > a_{\mathrm{pl}}$ across all tested codes. This indicates that the planet's orbital distance is consistently shorter than the maximal extent of its host star assuming single-star evolution.
}
\label{fig:3}
\end{figure}

\clearpage
\bibliography{maintext}
\newpage

\section*{Methods}
\subsection*{Orbital properties of the 8 UMi system}
\textbf{Radial velocity analysis}: The first radial velocity dataset for 8 UMi comprises measurements taken from the Bohyunsan Optical Astronomy Observatory (BOAO) using the using the Bohyunsan Observatory Echelle Spectrograph \citeSupp{Kim_2007} (BOES), which are available from the 8 UMi discovery paper \citeSupp{Lee2015}. We obtained additional high-precision radial velocity observations (Extended Data Table \ref{tab:rv}) using the HIRES spectrometer \citeSupp{Vogt1994} at the Keck-1 Telescope on Maunakea, Hawai`i. All data was reduced using the standard procedure of the California Planet Search \citeSupp{Howard_2010} (CPS). For all observations we placed an iodine-cell in the light path to project a series of fiducial absorption lines onto the stellar spectrum. These references allow to track the instrumental profile and precisely wavelength-calibrate the observed spectra. We also collected a high signal-to-noise iodine-free template spectrum, which together with the instrumental point-spread function (PSF) and iodine transmission function is a component of the forward model employed by the CPS Doppler analysis pipeline \citeSupp{Howard_2010, Butler_1996}.

We used the probabilistic modelling toolkit \texttt{exoplanet}\citeSupp{exoplanet, pymc3} version 0.5.4 to fit a Keplerian orbit to the radial velocity time series. Radial velocity offsets ($\gamma$) and jitter terms specific to BOAO ($\sigma_{\mathrm{BOAO}}$) and HIRES ($\sigma_{\mathrm{HIRES}}$) instruments were included as free parameters in the fit. Additionally, long-term trends in the radial velocity data were parameterized with a quadratic curve comprising a linear acceleration term ($\dot{\gamma}$) and a curvature term ($\ddot{\gamma}$). The priors for 
period $P$ and semi-amplitude $K$ were defined as normally distributed in log-space, such that $\log(P)\sim\mathcal{N}(\log(93.5), 10)$, and $\log(K)\sim\mathcal{N}(\log(56.5), 10)$. Those for $\omega$, $\dot{\gamma}$, and $\ddot{\gamma}$ were uniformly distributed, where
$\omega\sim\mathcal{U}$ ($-2\pi, +2\pi$), $\dot{\gamma}\sim\mathcal{U}(0, 0.1)$, and $\ddot{\gamma}\sim\mathcal{U}(0, 0.001)$.
The eccentricity ($e$) prior was described by Beta$\,(0.867, 3.03)$, following a distribution tailored towards empirical observations of extrasolar planet eccentricities \citeSupp{Kipping_2013}. The priors for $\gamma$ of each instrument were normally distributed, with $\gamma_{\mathrm{BOES}}\sim\mathcal{N}(-9.3, 200)$, $\gamma_{\mathrm{HIRES}}\sim\mathcal{N}(-2.4, 200)$. Meanwhile, those for $\sigma$ were parameterized with half-normal distributions, such that $\sigma_{\mathrm{BOES}}\sim\lvert\ \mathcal{N} \rvert (0, 100)$ and $\sigma_{\mathrm{HIRES}}\sim\lvert\ \mathcal{N} \rvert(0, 100)$.
The time of periastron passage, $t_P$ was fitted within the interval $0.5t_0 + \phi\times P$, where $\phi\sim\mathcal{U}(0,1)$ and $t_0$ is the midpoint in time between the first and last radial velocity observations. 

Following standard Markov chain Monte Carlo (MCMC) procedures,
\textit{maximum a posteriori} estimates of the Keplerian orbital parameters were determined through iterative fits to the time series, followed by the exploration of parameter uncertainties using the No-U-Turn Sampler \citeSupp{NUTS}. We performed 5,000 tuning iterations and 5,000 draws from two chains during the posterior sampling process, with random seeds of \texttt{39091} and \texttt{39095}. The process resulted in a Potential Scale Reduction Factor below 1.001 for each explored parameter, indicating posterior convergence. The sampled posterior distribution is presented in Extended Data Fig. \ref{fig:mcmc_corner}. The fitted jitter terms from BOES and HIRES are consistent with one another with values $\sim10$ m/s, which is expected from core-helium burning giants \citeSupp{Yu_2018, Tayar_2019}. \\

\noindent \textbf{Characterizing the outer companion}: An underlying long-term trend in the radial velocity data, as determined by non-zero values of first- ($\dot{\gamma}$) and second-order ($\ddot{\gamma}$) velocity derivatives, suggests the presence of an additional, outer companion to the system. These derivatives were defined with respect to a reference time, $t_{\mathrm{ref}}=2457508.0173445$. We compared radial velocity models with and without these additional terms, using
the Watanabe-Akaike Information Criterion \citeSupp{Watanabe_2010} (WAIC) metric. The score was measured as a deviance value, where smaller scores indicate better predictive accuracy.
We found the model including a trend (WAIC = 1265.80) was favoured over the one without (WAIC = 1282.37). The improvement in WAIC in using the trended model is significantly greater than the standard error in WAIC differences between the two models ($d_{\mathrm{WAIC}}=7.64$) and therefore the RV variations are better predicted with a trend, and suggest the presence of an outer companion with a longer period than our combined observational baseline of 12.5 years.

To constrain the mass and semi-major axis of the outer companion, we fitted a Keplerian orbit using joint constraints from the host star's long-term radial velocity trend and astrometric acceleration \citeSupp{Lubin_2022}. A broad range of masses, semi-major axes, $i$, $\omega$, $e$, and mean anomaly values were explored using MCMC procedures to identify orbits that fit the radial velocity derivatives ($\dot{\gamma}, \ddot{\gamma}$). To apply the astrometric constraints, we computed the expected proper motion vectors for each calculated orbit in the Hipparcos and 
\textit{Gaia} DR3 epochs, which were then contrasted against the observed change in proper motion for host star 8 UMi between the two epochs \citeSupp{Brandt_2021}. The results from separately and jointly fitting the radial velocity and astrometry constraints are shown in Extended Data Fig. \ref{fig:rv_astrometry}, where we conclude that the outer companion orbits at a distance greater than 5 au at a 67\% confidence level.

\subsection*{Stellar activity}
\indent Previously, the discovery of 8 UMi b was established using the lack of correlations between radial velocity measurements with H-$\alpha$ line variations, spectral line bisectors, and Hipparcos photometry \citeSupp{Lee2015}. We further established the planet's existence in this work using measured Ca II H and K spectral line variations and  spectropolarimetric observations of the host star to estimate its stellar activity level. We computed Ca II H and K indices ($S_{HK}$) following established calibration methods \citeSupp{Isaacson_2010} for Keck data. The results in Extended Data Fig. \ref{fig:rv_sval} indicate no significant correlation between 8 UMi's radial velocity measurements with $S_{HK}$ and no significant 93-day variation within its computed Generalized Lomb-Scargle (GLS) periodogram \citeSupp{Zechmeister_2009}.
The absence of strong chromospheric activity for 8 UMi is further evidenced in Extended Data Fig. \ref{fig:activity_absorp}a-b by its lack of strong Ca II H and K emission features. Extended Data Fig. \ref{fig:activity_absorp}c demonstrates the lithium-richness of 8 UMi at the 6707.8\r{A} Li I absorption line in comparison with the Li-normal \citeSupp{Liu_2014} core helium-burning giant $\mu$ Pegasi \citeSupp{polarbase}. We adopt a lithium abundance as $\mathrm{A(Li)}= 2.0\pm0.2$ for 8 UMi, which is the average and standard deviation of previously published \citeSupp{Kumar_2011_supp, Charbonnel_2020_supp} Local Thermodynamic Equilibrium (LTE) measurements of lithium for the star.

To estimate the strength of magnetic fields potentially present at the surface of the host star, we collected four spectropolarimetric observations between February 2022 and July 2022 with ESPaDOnS at the Canada-France-Hawaii Telescope (CFHT). The observed data comprised intensity ($I$), Stokes $V$, and null ($N$) polarisation spectra, reduced with the Libre-Esprit \citeSupp{Donati_1997} and Upena pipelines at CFHT. To detect the presence of Zeeman signatures within the spectra, we applied the improved Least Squares Deconvolution technique \citeSupp{Donati_1997, Kochukhov_2010} using line masks from the Vienna Atomic Line Database \citeSupp{Kupka_1999} corresponding to a model atmosphere of $T_{\mathrm{eff}}=4750$K and $\log(g)=2.5$ dex. All LSD profiles for 8 UMi in Extended Data Fig. \ref{fig:lsd_profiles} show null detection of Stokes V Zeeman signatures, which correspond to false alarm probabilities (FAP) \citeSupp{Donati_1997} $>1\times10^{-3}$. This firmly excludes the presence of a $\sim5\,$G mean longitudinal magnetic field strength expected for an active red giant rotating at a period of 90 days \citeSupp{Auriere_2015}. 

These observations are benchmarked against the known active red giant TYC 3542-1885-1 (KIC 8879518). This active star has a measured \textit{Kepler} photometric rotation period of $109$ days \citeSupp{Gaulme_2020}, and was chosen as a control target by virtue of having a similar evolutionary state, mass ($M=1.61\,M_{\odot}$) and radius ($R=10.89\,R_{\odot}$) to the host star 8 UMi. Additionally, TYC 3542-1885-1 is also lithium-rich \citeSupp{Wheeler_2021}, albeit with higher abundance levels to 8 UMi (Extended Data Fig. \ref{fig:activity_absorp}c). From our Keck/HIRES observations of TYC 3542-1885-1 between 2 April 2022 and 1 August 2022 (Extended Data Fig. \ref{fig:rv_sval}), the star exhibits a quasi-periodic $\sim65\,$d radial velocity variation that correlates strongly with its measured $S_{HK}$, inferred to be caused by a secondary starspot feature.
Despite lower SNR spectropolarimetric data, ESPaDOnS observations of TYC 3542-1885-1 show marginal ($1\times10^{-5}<\mathrm{FAP}\leq1\times10^{-3}$) evidence of a mean longitudinal magnetic field (Extended Data Fig. \ref{fig:lsd_profiles}) at strengths of $\sim3\,$G, which is broadly consistent with the expected field strength for a 100-200 d rotating red giant \citeSupp{Auriere_2015}. From our analysis of the control target, we expect the host star 8 UMi to show similar patterns in chromospheric and surface magnetic activity levels if its 93-day radial velocity variations are stellar in nature, but such activity signatures were not observed.

Extended Data Fig. \ref{fig:phot_var} shows our analysis of 8 UMi's photometry from different sources. A starspot mimicking 8 UMi's $\sim50\,$m/s radial velocity variations would show photometric variability at the $2.5-3$\% level, based on simulated results of a generic starspot (e.g., 600$\,$K spot contrast, 3\% visible hemisphere coverage, $60^{\circ}$ longitude) using the Spot Oscillation And Planet 2.0 tool \citeSupp{Dumusque_2014}. We found 8 UMi's time series from Hipparcos \citeSupp{Hipparcos} and from ASAS-SN \citeSupp{Shappee_2014, Kochanek_2017} 3-pixel ($\sim24$ arcsec) radius aperture photometry to be constant to within $0.5\%$ and showing no significant 93-day periodic variation within their computed GLS periodograms. Additionally, we examined TESS short-cadence Simple Aperture Photometry light curves corrected using the TESS Systematics-Insensitive Periodogram (TESS-SIP) \citeSupp{Hedges_2020}, which confirmed the same null result from a search of periods spanning 20-200 days. We thus conclude that 8 UMi's radial velocity variations are planetary in nature.

\subsection*{Asteroseismic measurements}
We used all TESS light curves for 8 UMi across Cycles 2 and 4 (total of 12 Sectors) that were produced by the TESS Science Operations Center \citeSupp{Jenkins_2016} (SPOC), which are available at the Mikulski Archive for Space Telescopes (MAST) as of January 2023. We determined the oscillation frequency at maximum power, $\nu_{\mathrm{max}}=45.29\pm0.43\,\mu$Hz, by fitting the light curve's power density spectrum with a model comprising a Gaussian oscillation power envelope superimposed upon a background of three granulation parameters and white noise \citeSupp{Themessl_2020}. We identified oscillation modes using a well-tested automated peak detection method \citeSupp{Montellano_2018} to fit Lorentzian-like models to the resonant peaks within the power envelope. We determined the large frequency spacing, $\Delta\nu=4.80\pm0.01\,\mu$Hz, by fitting the detected radial modes to the red giant oscillation pattern \citeSupp{Mosser_2011} following the asymptotic relation of stellar acoustic modes \citeSupp{Tassoul_1980}. 

We measured the dipole mode period spacing, $\Delta\Pi$, by converting the frequency axis of the oscillation spectrum into stretched periods \citeSupp{Mosser_2015}. Following established automated procedures \citeSupp{Vrard_2016_supp}, stretched period spectra were computed over trial values of $\Delta\Pi$ and mixed mode coupling factors ($q_c$), with the best solution providing the highest power in the Fourier transform of the resulting stretched period spectrum. 
By searching between $\Delta\Pi$ of 40-400$\,$s and $q_c$ of $0.05-0.80$, we determined the solution at $(\Delta\Pi ,q_c)=(320\,$s$, 0.35)$ as the global maximum, which firmly identifies 8 UMi as a core helium-burning giant \citeSupp{Mosser_2017}.

\subsection*{Stellar modelling}
Four different codes \citeSupp{Rodrigues_2014, Rodrigues_2017, Yildiz_2016, Jiang_2021, Silva-Aguirre_2022} were used to model host star 8 UMi and estimate its fundamental stellar properties. We constrained the modelling to core helium-burning stars and used as inputs asteroseismic parameters $\Delta\nu$ and $\nu_{\mathrm{max}}$, along with atmospheric parameters $T_{\mathrm{eff}}=4847\pm100\,$K and [M/H]$=-0.03\pm0.10\,$ dex. These were adopted from the discovery paper \citeSupp{Lee2015}, with the uncertainties (standard deviation) inflated from their original published values \citeSupp{Tayar_2022}. We report the best-fitting estimates across the modelling codes, which are $M=1.51\pm0.05$ (stat) $\pm0.04$ (sys) $\,M_{\odot}$ , $R=10.73\pm0.11$ (stat) $\pm0.08$ (sys) $\,R_{\odot}$, $L=52.9\pm4.9$ (stat) $\pm3.3\,L_{\odot}$ (sys).  The central values were provided by one of the codes \citeSupp{Yildiz_2016} that showed the smallest difference to the
median derived mass, with systematic uncertainties as the standard deviation of an estimated parameter over
all codes. By combining errors in quadrature, these results translate to average uncertainties of $0.06\,M_{\odot}$ for mass, $0.14\,R_{\odot}$ for radius, and $5.9\,L_{\odot}$ for luminosity.

We verified our modelled results using independent estimation methods, where for each the 1$\sigma$ uncertainty is reported as that method's standard deviation. Using model-calibrated asteroseismic scaling relations \citeSupp{Sharma_2016, Stello_2022}, we measured $M_{\mathrm{scal}}=1.54\pm0.06\,M_{\odot}$ and $R_{\mathrm{scal}}=10.72\pm0.16\,R_{\odot}$. Additionally, we fitted Kurucz stellar atmosphere models to the broadband (0.2--22~$\mu$m) spectral energy distribution (SED) of the host star \citeSupp{Stassun:2016,Stassun:2017,Stassun:2018} (see Extended Data Fig. \ref{fig:sed}), including extinction based on the maximum line-of-sight value from published Galactic dust maps \citeSupp{Schlegel:1998}. Integration of the unreddened model SED, combined with a \textit{Gaia} DR3 parallax $\varpi=6.128\,$mas (with no systematic offset applied \citeSupp{StassunTorres:2021}), resulted in $L_{\mathrm{SED}}=53.81\pm1.84\,L_{\odot}$ and subsequently $R_{\mathrm{SED}}=10.19\pm0.36\,R_{\odot}$, which are in agreement to within 1$\sigma$ with our modelled estimates.

We determined theoretical estimates of the maximum size of a pre-core helium-burning star by interpolating the radius of its stellar envelope at the tip of the red giant branch ($R_{\mathrm{tRGB}}$) across mass and metallicity using various \citeSupp{Choi_2016, Hidalgo_2018, Dotter_2008, Tayar_2022, Bressan_2012, Weiss_2008} isochrone grids, all of which are based on canonical single-star evolution. The interpolation produces a boundary above which the current semi-major axis of 8 UMi b would be smaller than $R_{\mathrm{tRGB}}$ of its host star (see main text Fig. \ref{fig:3}). We measured the closest extent of each isochrone's boundary to 8 UMi's mass and metallicity and found with a 3-8$\sigma$ confidence that 8 UMi b, had it formed in-situ at 0.46 AU, would have been within its host star's stellar envelope during the giant branch phase assuming canonical stellar evolution. 



\subsection*{Exploring survival scenarios}


\noindent \nzradd{\textbf{Tidal migration and circularization}:  Assuming a constant quality factor $Q$, the tidal migration timescale $\tau_{\mathrm{tide}}$ is roughly the evolutionary timescale of the semi-major axis due to tides:}
\begin{equation}
    \left\lvert\frac{\dot{e}}{e}\right\rvert_{\mathrm{tide}} \sim \left\lvert\frac{\dot{r}}{r}\right\rvert_{\mathrm{tide}}  = \frac{1}{\tau_{\mathrm{tide}}} = \frac{1}{\tau^\star_{\mathrm{tide}}} + \frac{1}{\tau^p_{\mathrm{tide}}} \simeq \frac{1}{\mathrm{min}(\tau^\star_{\mathrm{tide}},\tau^p_{\mathrm{tide}})}
\end{equation}

\noindent \nzradd{where $\tau^\star_{\mathrm{tide}}$ and $\tau^p_{\mathrm{tide}}$ account for tidal dissipation in the star and planet, respectively.
These timescales are estimated as\citeSupp{goldreich1966q}}


\begin{subequations}
    \begin{gather}
        \tau^\star_{\mathrm{tide}} \simeq \frac{1}{9\pi}Q_\star\frac{M_\star}{M_p}\left(\frac{a}{R_\star}\right)^5P \approx 60\,\mathrm{Gyr}\left(\frac{Q_\star}{10^5}\right)\left(\frac{M_\star}{1.51\,M_\odot}\right)^{1/2}\left(\frac{M_p}{1.65\,M_J}\right)^{-1}\left(\frac{a}{0.46\,\mathrm{au}}\right)^{13/2}\left(\frac{R_\star}{10.73\,R_\odot}\right)^{-5} \\
        \tau^p_{\mathrm{tide}} \simeq \frac{1}{9\pi}Q_p\frac{M_p}{M_\star}\left(\frac{a}{R_p}\right)^5P \approx 10^5\,\mathrm{Gyr}\left(\frac{Q_p}{10^5}\right)\left(\frac{M_p}{1.65\,M_J}\right)\left(\frac{M_\star}{1.51\,M_\odot}\right)^{-3/2}\left(\frac{a}{0.46\,\mathrm{au}}\right)^{13/2}\left(\frac{R_p}{1.50\,R_J}\right)^{-5}
    \end{gather}
\end{subequations}

\noindent \nzradd{where the star and planet tidal quality factors $Q_\star$ and $Q_p$ are scaled to the lower end of $Q$ values reported in literature for Jovian planets\citeSupp{wu2005origin} and red giants\citeSupp{essick2015orbital} with $Q_\star\sim Q_p\gtrsim10^5$, and assuming a representative Jovian planet radius \citeSupp{Fortney_2021} of $1.5 R_J$.
Hence, present-day tidal circularization and migration takes place on timescales characteristically larger than a Hubble time, and can be ignored.}

\nzradd{Of course, during the red giant's expansion prior to the helium flash, its radius would have greatly exceeded its current-day value, significantly enhancing the tidal dissipation and dramatically shortening the orbit's circularization timescale.
If, during this expansion, the outer radius of the star barely grazed the planet's orbit at its helium-flash radius $a\approx R_\star\approx0.7\,\mathrm{au}$, a similar calculation yields a minimal dissipation timescale  $\tau^{\star,\mathrm{flash}}_{\mathrm{tide}} \simeq 1.6\,\mathrm{Myr}$.}
\nzradd{This approximates a lower bound to the planet's migration timescale in the most extreme scenario where a very fine-tuned dynamical process brings the planet as close to the star as possible during its helium flash.
This timescale should be compared to the timescale of $\sim50\,\mathrm{kyr}$ on which the star shrinks following the initial helium flash, when its radius is largest.
Hence, the planet could not migrate inwards as fast as the star contracts following the helium flash. Even under these fine-tuned scenarios, tidal circularization and migration cannot act fast enough to deliver the planet to its current orbital separation and eccentricity.}\\

\noindent \textbf{8 UMi b mass constraints}: Assuming orbits are randomly oriented, the probability of an orbit having inclination $i$ is proportional to $\sin i$. Given 8 UMi b's mass of $M_p \sin i = 1.65\,M_J$, the planet's orbit would need to be nearly face-on ($i \leq 7.3^{\circ}$) for its mass to exceed $13\,M_J$, which is a threshold adopted as the deuterium-burning limit for brown dwarfs \citeSupp{Spiegel_2011}. This range of $i$ occurs with a probability of 0.8\%, based on draws from an isotropic distribution.

Host star 8 UMi has low astrometric excess noise as reported by the \textit{Gaia} space mission. This noise is quantified by the Re-normalized Unit Weight Error (RUWE), and has a value of $\rho_{\mathrm{DR2}}=1.09$ for \textit{Gaia} Data Release 2 (DR2) during which 17 visibility periods were reported over the observing baseline. The angular perturbation to the single source of the photocentre as fitted by \textit{Gaia} can be estimated \citeSupp{Belokurov_2020} as $\delta{\theta}\approx 0.53\sigma_{\varpi}\sqrt{N(\rho_{\mathrm{DR2}} - 1)} = 0.076\,$mas , where $\sigma_{\varpi}$ is the star's parallax error of 0.024 mas and $N=192$ is the number of good along-scan \textit{Gaia} observations. Given the near-zero eccentricity constraint from the radial velocity data, we considered perturbations in the limit of a face-on alignment for a circular orbit \citeSupp{Penoyre_2020}, such that
\begin{equation}
   \delta{\theta} = \frac{\varpi a |q-l_r|}{(1+q)(1+l_r)},
\end{equation}
where the DR2 parallax $\varpi=6.123\,$mas, the semi-major axis $a=0.462\,$au, $q = M_{\mathrm{planet}}/M_{\star}$ is the mass ratio, and $l_r= L_{\mathrm{planet}}/L_{\star}$ is the luminosity ratio, which we approximate as zero. 
Assuming the perturbation comes purely from the orbiting companion, $q$ is predicted to be $\sim0.028$ to yield a sufficiently large $\delta{\theta}$ that produces the observed $\rho_{\mathrm{DR2}}$. This suggests that companion masses above $0.043M_{\odot}$ or $45\,M_J$ would induce larger astrometric noise excesses than as measured by $\rho_{\mathrm{DR2}}$ and therefore rules out 8 UMi b as a stellar companion. In \textit{Gaia} Data Release 3 (DR3), 8 UMi is reported to have a high astrometric fidelity \citeSupp{Rybizki_2022} (value = 1), with a nearly identical parallax of $\varpi_{\mathrm{DR3}}=6.128\,$mas and $\rho_{\mathrm{DR3}} = 0.80$ based on 23 visibility periods. This value of $\rho_{\mathrm{DR3}}$, when considered in tandem with $\rho_{\mathrm{DR2}}$, is again inconsistent with excess astrometric scatter from a close stellar companion \citeSupp{Penoyre_2022}.
\\


\noindent \textbf{Common-envelope evolution}: \nzradd{
The observed orbital separation $\approx0.5\,\mathrm{au}$ eliminates the possibility of 8 UMi b initiating a successful common envelope, independent of its mass.
Assuming that 8 UMi b starts at an orbital separation $a_i\approx0.7\,\mathrm{au}$ (the tip of the red giant branch), the orbital energy available to unbind the envelope is}

\begin{equation}
    \Delta E_{\mathrm{orb}} \simeq \frac{GM_pM_c}{2a_f} - \frac{GM_pM_\star}{2a_i} = \frac{1}{2}GM_p\left(\frac{M_c}{a_f} - \frac{M_*}{a_i}\right)
\end{equation}

\noindent \nzradd{where $M_*\approx1.5\,M_\odot$, $M_c\approx0.25\,M_\odot$, and $a_f\approx0.5\,\mathrm{au}$.
However, because $M_c/M_*<a_f/a_i$, we find that $\Delta E_{\mathrm{orb}}<0$, which implies that the planet ends up \textit{less} bound than before. The observed semi-major axis of 8 UMi b therefore precludes the orbit from having shrunk enough to unbind any of the envelope, regardless of the value of $M_p$.
Even in the absence of a common-envelope event, the companion would almost certainly not survive temporary engulfment by the host star.
The maximum inspiral timescale $t_{\mathrm{insp}}$ for a $0.1\,M_\odot$ red dwarf in the envelope of 8 UMi can be estimated as the time for the companion's orbital energy to be reduced by ram pressure \citeSupp{Macleod_2018_supp}. For a drag force $F_D$ and orbital velocity $v$, the inspiral time is $t_{\mathrm{insp}}\sim E_{\mathrm{orb}}/F_D v\propto M_p/R_p^2\sim10^2 - 10^4\,\mathrm{yr}$. Assuming that the host star had previously engulfed the planet at its current orbital distance of $0.5\,$au, the calculated inspiral timescale is much smaller than the $\sim2\,\mathrm{Myr}$ required for the host star to reach $R_{\mathrm{tRGB}}$ as estimated by stellar models \citeSupp{Choi_2016} . 
}
\subsection*{Modelling the binary merger scenario} \label{binaryscenario}

Using Modules for Experiments in Stellar Astrophysics (\textsc{mesa}, version r22.05.1\citeSupp{paxton2010modules,paxton2013modules,paxton2015modules,paxton2018modules,paxton2019modules}), we simulated coupled orbit---stellar binary evolution scenarios resulting in a merger of a red giant with a helium white dwarf companion as a pathway for 8 UMi b's survival. These scenarios aimed to simulate a first, stable mass transfer phase that stably strips the primary followed by a second, unstable mass transfer phase that merges the two stars. Leading up to the first mass transfer phase, we simulated the stellar evolution and orbits of both stars together until the primary has been totally stripped into a helium white dwarf. After the first mass transfer phase, we modelled the primary purely as a point mass with the mass of the primary's stripped core while still evolving the orbits of both stars. As observational constraints, 
the simulated initial component masses $M_1$ and $M_2$ were set such that the final mass of the remnant, $M$, is 8 UMi's observed stellar mass of $\approx1.5\,M_\odot$. Additionally, to avoid dynamically destabilizing the planet, the semi-major axis of the simulated binary throughout its evolution cannot exceed $\gtrsim30\%$ of the planet's own semi-major axis\citeSupp{holman1999long}\nzradd{, which evolves to conserve orbital angular momentum}. 

We included magnetic braking and tidal synchronization in our modelling, where we assume that tides instantaneously synchronize the stellar spins to the orbital frequency, which is a good approximation at short orbital periods.
The magnetic braking torque is a strong function of spin frequency, with $\dot{J}\propto\Omega^3$ in standard magnetic braking prescriptions\citeSupp{rappaport1983new}, resulting in the high sensitivity of the simulation outcomes to the initial orbital separation. Additionally,  we assumed mass transfer in the form of fast wind off of the accretor. This form of mass loss is governed by the $\beta$ parameter\citeSupp{tauris2006formation}, with $\beta=0$ indicating that no mass is lost from the system and $\beta=1$ indicating that all mass is lost and therefore no mass is transferred to the accretor. 
Assuming no mass is lost during the merger, the final mass of the remnant is given by
\begin{equation} \label{eqn:mass_transfer}
    M = (1-\beta)(M_1-M_c) + M_c + M_2
\end{equation}
\noindent where $M_c$ is the maximum mass of the primary's core (i.e., the mass of the stripped helium white dwarf). We determined $M_1$ and $M_2$ by specifying $\beta$ and a mass ratio $q$, assuming $M\approx1.50\,M_\odot$ and $M_c\approx0.25\,M_\odot$. A fully conservative mass transfer ($\beta=0$) was found to result in an unstable first mass transfer phase, while fully non-conservative mass transfer ($\beta=1$) destabilized the planet through excessive orbital expansion. We thus used a fiducial choice of parameters $q=0.7$ and $\beta=0.6$ ($M_1=1.23\,M_\odot$, $M_2=0.86\,M_\odot$),
and subsequently identified that models with initial orbital periods $1.7\,\mathrm{d}\lesssim P_{\mathrm{init}}\lesssim2.3\,\mathrm{d}$ can stably strip the envelope of the primary completely, with lower period models either evolving into contact binaries (or merging prematurely) and higher period models undergoing unstable mass transfer immediately. These initial orbital periods and the corresponding fiducial parameters are values that have been observed in solar-type binaries \citeSupp{raghavan2010survey}, with previously reported statistics \citeSupp{Moe_2017} estimating the frequency of companions per decade of orbital period of short-period ($\simeq 3\,$days) binary systems with $q>0.3$ to be $0.017\pm0.007$.

Extended Data Fig. \ref{fig:time_plot} shows an example fiducial model with $P_{\mathrm{init}}=2\,\mathrm{d}$.
Magnetic braking on the main sequence reduces the orbital separation by a factor of $\sim2$ until the onset of the first stable mass transfer phase, which is when the donor reaches the red giant branch. During this time ($t\sim4.2$--$5.6$ Gyr in Extended Data Fig. \ref{fig:time_plot}) the envelope of the primary is stripped, the secondary accretes mass, and the binary's semi-major axis increases to $\sim1.7$ times its initial value, but not enough to destabilize the planet's orbit. \nzradd{The resulting binary system forms an EL CVn-like binary with a period shorter than 10 days.

The \textsc{mesa} simulations of Chen et al.\citeSupp{Chen_2017} predicted that a range of $1.9\,$d $\lesssim P_{\mathrm{init}}\lesssim2.4\,$d and $0.67 \lesssim q \lesssim 1$ can yield donors that undergo mass transfer at the base of the red giant branch for $1.1M_{\odot} < M_1 < 1.25M_{\odot}$ ($\beta=0.5$ has been assumed in this case), which is consistent with our fiducial model.
They additionally performed a population synthesis study and estimated a steady production rate of EL CVn binaries $\approx0.015\,\mathrm{yr}^{-1}$ over the last few $\mathrm{Gyr}$, assuming a Miller \& Scalo initial mass function in the range $0.08$--$100$ $M_\odot$\citeSupp{miller1979initial,eggleton1989distribution} and a constant star formation rate $5\,M_\odot\,\mathrm{yr}^{-1}$. We calculate such a model to yield a $\approx0.6\,\mathrm{yr}^{-1}$ production rate of $1$--$2\,M_\odot$ stars, which implies that $\sim2$--$3\%$ of red clump stars may have formed from merging EL CVn-like binaries.

This number is roughly comparable to the fraction of clump stars which are lithium-rich ($\sim1\%$\citeSupp{sneden1984search,yan2018nature,Zhang_2020_supp}) or R-type carbon stars ($\sim0.1\%$\citeSupp{izzard2007origin}), which are thought to form from similar merger processes.
It is also broadly consistent with the rate of merger remnants\citeSupp{rui2021asteroseismic} inferred for red giants from transient statistics\citeSupp{kochanek2014stellar}, close binary surveys\citeSupp{raghavan2010survey}, and red giant binarity\citeSupp{price2020close}.} When applying the inferred EL CVn-like binary merger fraction to the $>140$ substellar companions detected around red giants \citeSupp{RCdatabase}, roughly 1 in $\sim10^2$ companions may indeed orbit helium white dwarf--red giant merger remnants if we assume that an order-unity fraction of these exoplanet hosts are core helium-burning. Therefore, the hypothesis that 8 UMi is such a remnant is plausible.

Unstable mass transfer begins at the end of the simulation using our fiducial model when both the secondary's core mass and the helium white dwarf's mass are $\approx\!0.25 \, M_\odot$, such that a stellar merger results in a helium core mass of $\approx 0.5 \, M_\odot$, similar to ordinary red clump stars.
To determine if the second, unstable mass transfer phase results in a merger, we estimated the final semi-major axis $a_f$ of the binary from its initial semi-major axis $a_i$ by assuming that some fixed fraction $\alpha_{\mathrm{CE}}$
of the change in orbital energy is used to move the envelope to infinity\citeSupp{ivanova2013common}:
\begin{equation} \label{cee}
    |E_{\mathrm{bind}}| \sim \alpha_{\mathrm{CE}}\left(\frac{GM_{1c}M_{2c}}{2a_f} - \frac{GM_{1c}M_{2i}}{2a_i}\right),
\end{equation}
\noindent where $M_{1c}$ is the mass of the stripped helium white dwarf, and $M_{2i}$ and $M_{2c}$ are the total and core masses of the red giant prior to the merger event.
We calculated the binding energy of the envelope directly from our stellar models, including internal (thermal) energy and recombination energy, as
\begin{equation}
    |E_{\mathrm{bind}}| = \int^M_{M_{2c}}\left[u(m)+\epsilon(m)\right]\mathrm{d}m \, .
\end{equation}
\noindent Here, \nzradd{$u(m)$ and $\epsilon(m)$ are the specific potential and internal energies of the red giant, respectively, and}
we defined the boundary of the helium core to be where the hydrogen fraction falls below $1\%$. We used $\alpha_{\mathrm{CE}}=1/3$, which is consistent with observed helium white dwarf binaries\citeSupp{scherbakCE}. A stellar merger event is predicted when the post-common-envelope orbital separation is sufficiently small such that the helium core \nzradd{(of radius }$R_{2c}$\nzradd{)} of the giant is disrupted by the inspiraling helium white dwarf.
This occurs at an orbital separation $a_*=R_{2c}/f(q_f)$, where $q_f$ is the ratio of the masses of the red giant core and helium white dwarf, and $f(q_f)$ is an order-unity function given by Eggleton's formula\citeSupp{eggleton1983approximations} ($f(q_f)\approx0.38$ for equal-mass binaries). As demonstrated in Extended Data Fig. \ref{fig:magnetic_braking}, all our models
with $P_{\mathrm{init}} \lesssim 2.3 \, {\rm d}$ were predicted to attain final orbital separations $a_f<a_*$ and thus resulted in stellar mergers. Our mergers were assumed to occur with no mass loss, which we justify with the observed low eccentricity of 8 UMi b. Given a planetary post-mass-loss eccentricity of $e=\Delta M/M$, and assuming
an initially circular orbit with instantaneous mass loss, the star mass loss $\Delta M$ cannot be more than $\sim\!0.1 \, M_\odot$ in a time shorter than the planet's orbital period if the planet's nearly circular orbit is to be maintained.

\section*{Data availability}
TESS light curves processed by the SPOC pipeline are available from MAST 
(\url{https://archive.stsci.edu/}). The spectra for $\mu\,$Pegasi is accessible at \url{http://polarbase.irap.omp.eu/}.
Astrometric measurements for 8 UMi are openly available from the \textit{Gaia} archive (\url{https://gea.esac.esa.int/archive/}). The HIRES radial velocity measurements, ESPaDOnS spectra and spectropolarimetric data products, ASAS-SN time series, traces of the MCMC sampling from the radial velocity fits, MESA binary simulation inlists, and SED data are available at 
\url{https://zenodo.org/record/7668534}.

\section*{Code availability}
The radial velocity fitting was performed using the \texttt{exoplanet} code (\url{docs.exoplanet.codes/}).
The Generalized Lomb-Scargle periodogram implementation is available at (\url{https://github.com/mzechmeister/GLS}). TESS-SIP for correcting TESS systematics is provided at (\url{https://github.com/christinahedges/TESS-SIP}).
The asteroseismic modelling was performed using BASTA (\url{https://github.com/BASTAcode/BASTA}), the PARAM web tool (\url{http://stev.oapd.inaf.it/cgi-bin/param}), MESA (\url{https://docs.mesastar.org}).
The binary module of MESA was used for binary simulations.
Calibrated asteroseismic scaling relations used \texttt{asfgrid} (\url{http://www.physics.usyd.edu.au/k2gap/Asfgrid/}). Grids of isochrones publicly available are MIST (\url{https://waps.cfa.harvard.edu/MIST/}), PARSEC (\url{https://github.com/philrosenfield/padova\_tracks/releases/tag/v2.0}), Dartmouth and GARSTEC (\url{https://zenodo.org/record/6597404}), and BASTI (\url{http://albione.oa-teramo.inaf.it/}).  


\clearpage

\clearpage
\setcounter{figure}{0}
\renewcommand{\figurename}{Extended Data Figure}

\begin{figure}[ht]
\centering
\includegraphics[width=1\linewidth]{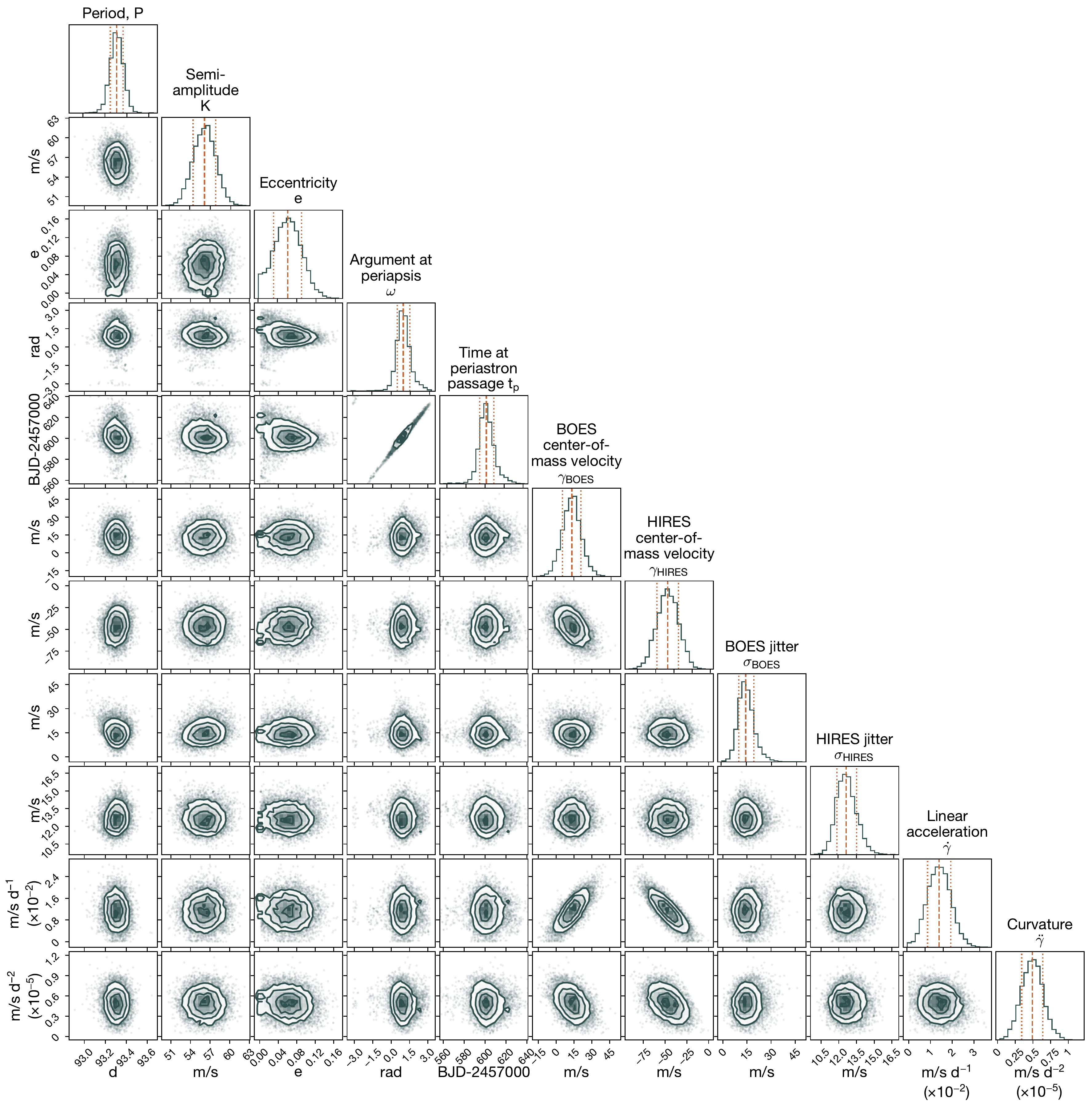}
\caption{\textbf{Posterior probability distribution of the fit to the combined BOES/HIRES radial velocity data.} The 16$^{\mathrm{th}}$, 50$^{\mathrm{th}}$, and 84$^{\mathrm{th}}$ percentile values of each fitted parameter are indicated with dashed lines.}
\label{fig:mcmc_corner}
\end{figure}

\begin{figure}[ht]
\centering
\includegraphics[width=0.8\linewidth]{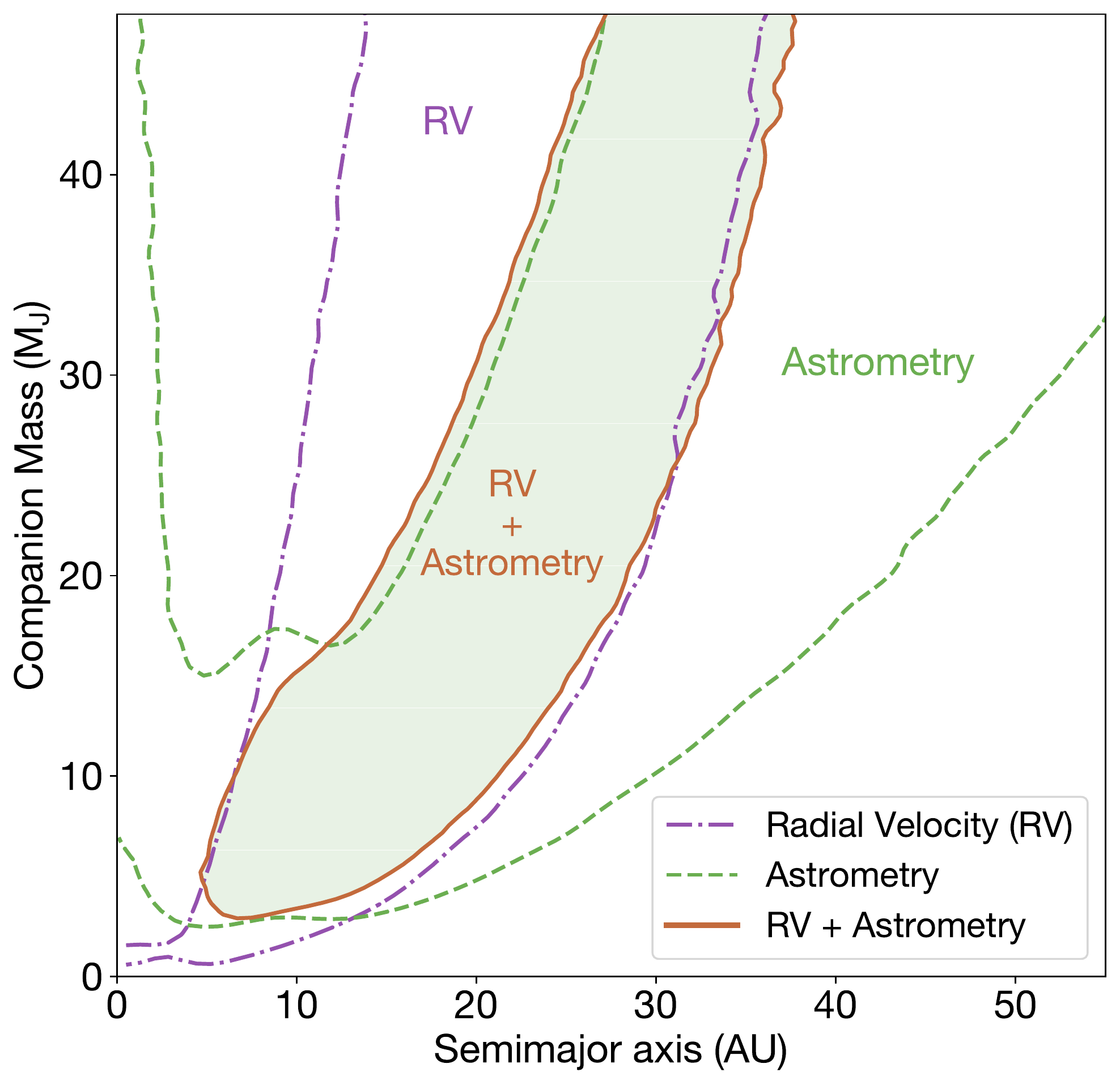}
\caption{\textbf{Constraints to the outer companion in the 8 UMi planetary system.} Contours indicate 67\% highest density intervals of permissible mass and separation values of the outer companion as estimated from the residuals of radial velocity measurements in Figure \ref{fig:1} (\textit{purple}), and from measurements of 8 UMi's \textit{Gaia} DR3-Hipparcos astrometric acceleration (\textit{green}). These two measurements jointly constrain the outer companion's mass and separation, which corresponds to the locus indicated by the shaded region between the contours.}
\label{fig:rv_astrometry}
\end{figure}

\begin{figure}[ht]
\centering
\includegraphics[width=1\linewidth]{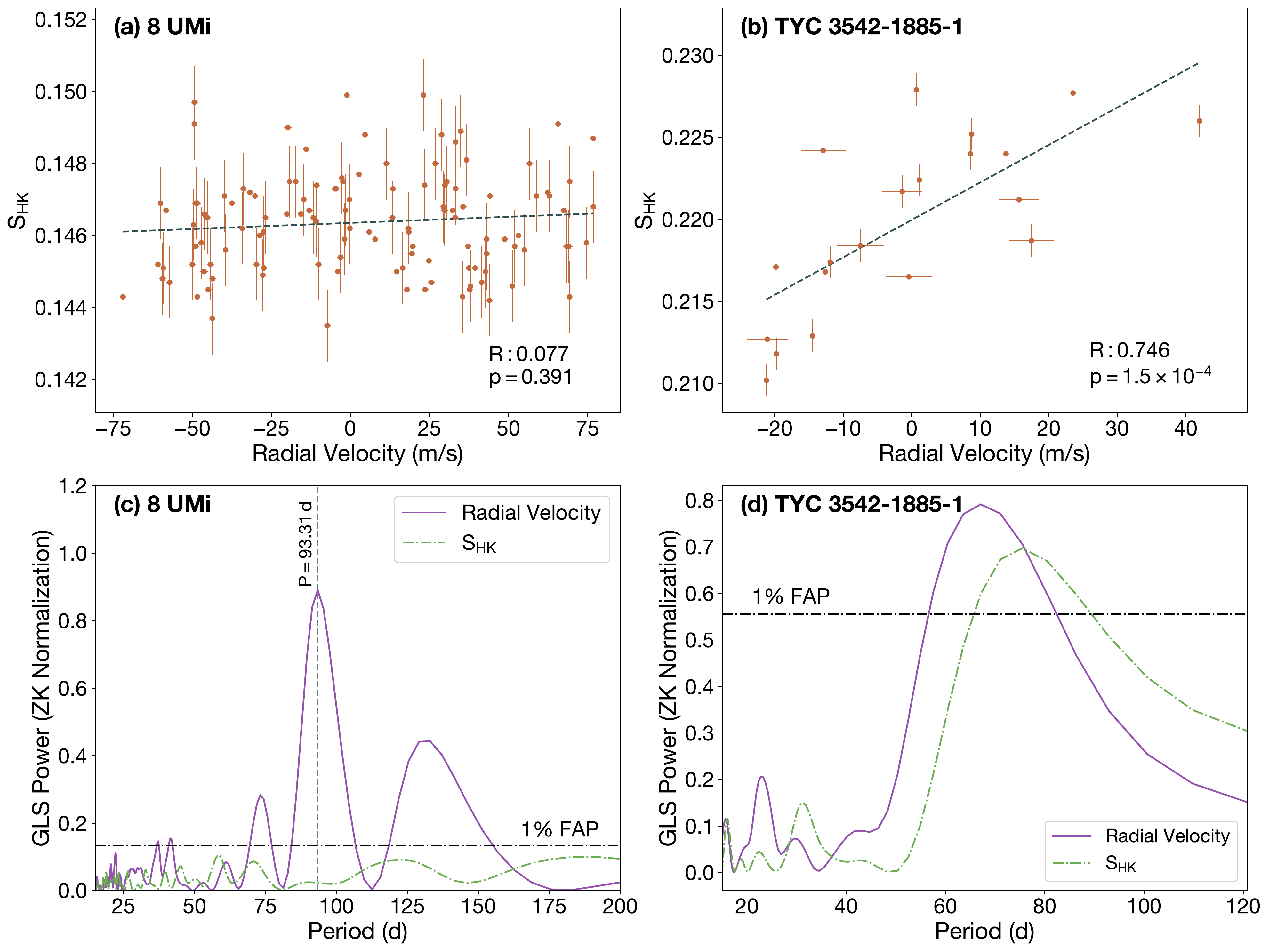}
\caption{\textbf{Stellar activity of the host star 8 UMi and active red giant TYC 3542-1885-1.} The chromospheric activity of both stars are estimated using 
Ca II H and K indices ($S_{HK}$) computed from Keck/HIRES spectra, with error bars indicating 1$\sigma$ (standard deviation) uncertainties. (\textit{a-b}) Variations of $S_{HK}$ with radial velocity from each star. Included for each are the Spearman correlation factors ($R$) and two-sided p-values ($p$) for the test whose null hypothesis is that $S_{HK}$ and radial velocity are uncorrelated. (\textit{c-d}) Generalized Lomb Scargle (GLS) periodograms of radial velocity measurements and $S_{HK}$. The vertical dashed line indicates 8 UMi b's orbital period, and the horizontal lines indicate the periodogram's False Alarm Probability (FAP).}
\label{fig:rv_sval}
\end{figure}

\begin{figure}[ht]
\centering
\includegraphics[width=1\linewidth]{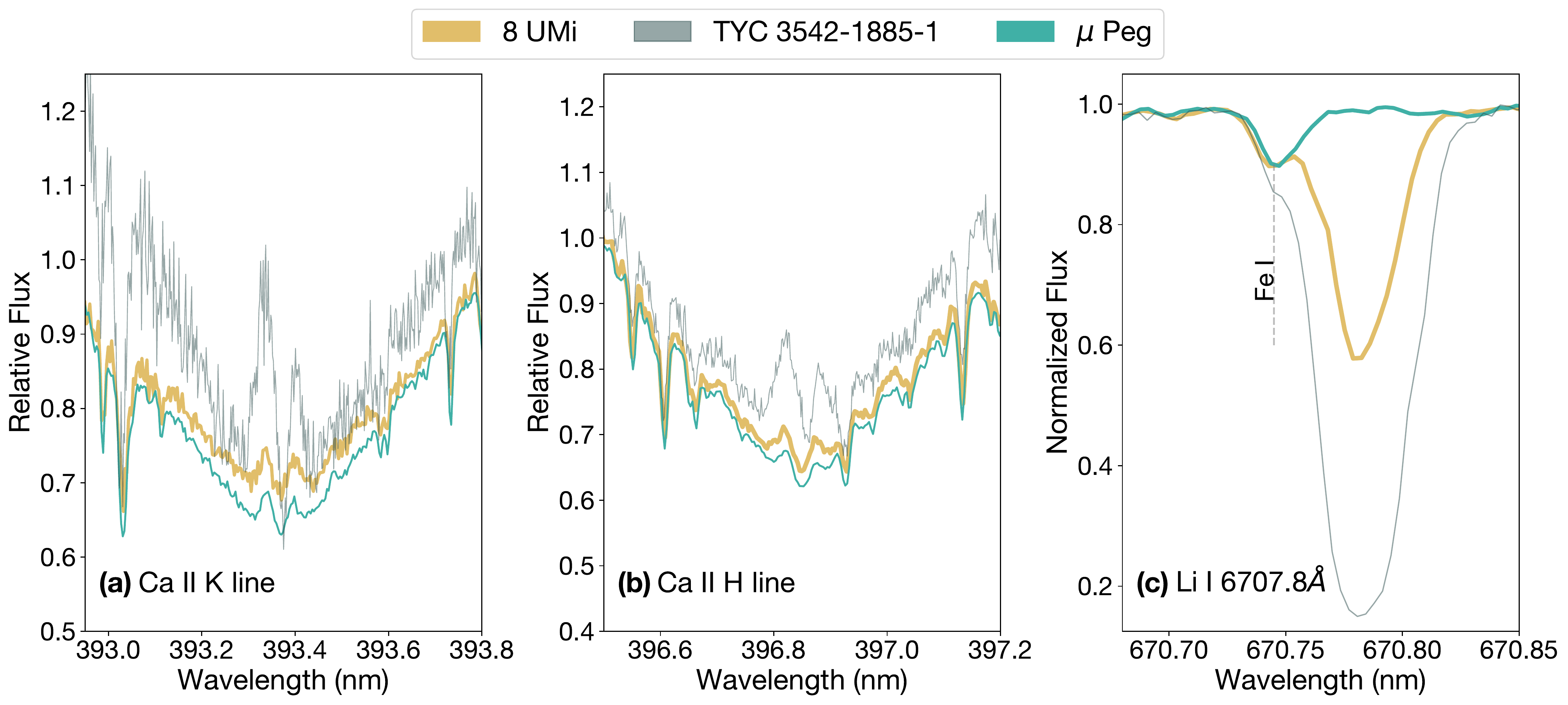}
\caption{\textbf{Spectral features of 8 UMi and active red giant TYC 3542-1885-1.}. Comparisons are additionally made with the inactive, Li-normal giant $\mu$ Pegasi. (\textit{a-b}) The Ca II H and K absorption lines. (\textit{c}) The 6707.8\r{A} Li I absorption line. }
\label{fig:activity_absorp}
\end{figure}

\begin{figure}[ht]
\centering
\includegraphics[width=0.8\linewidth]{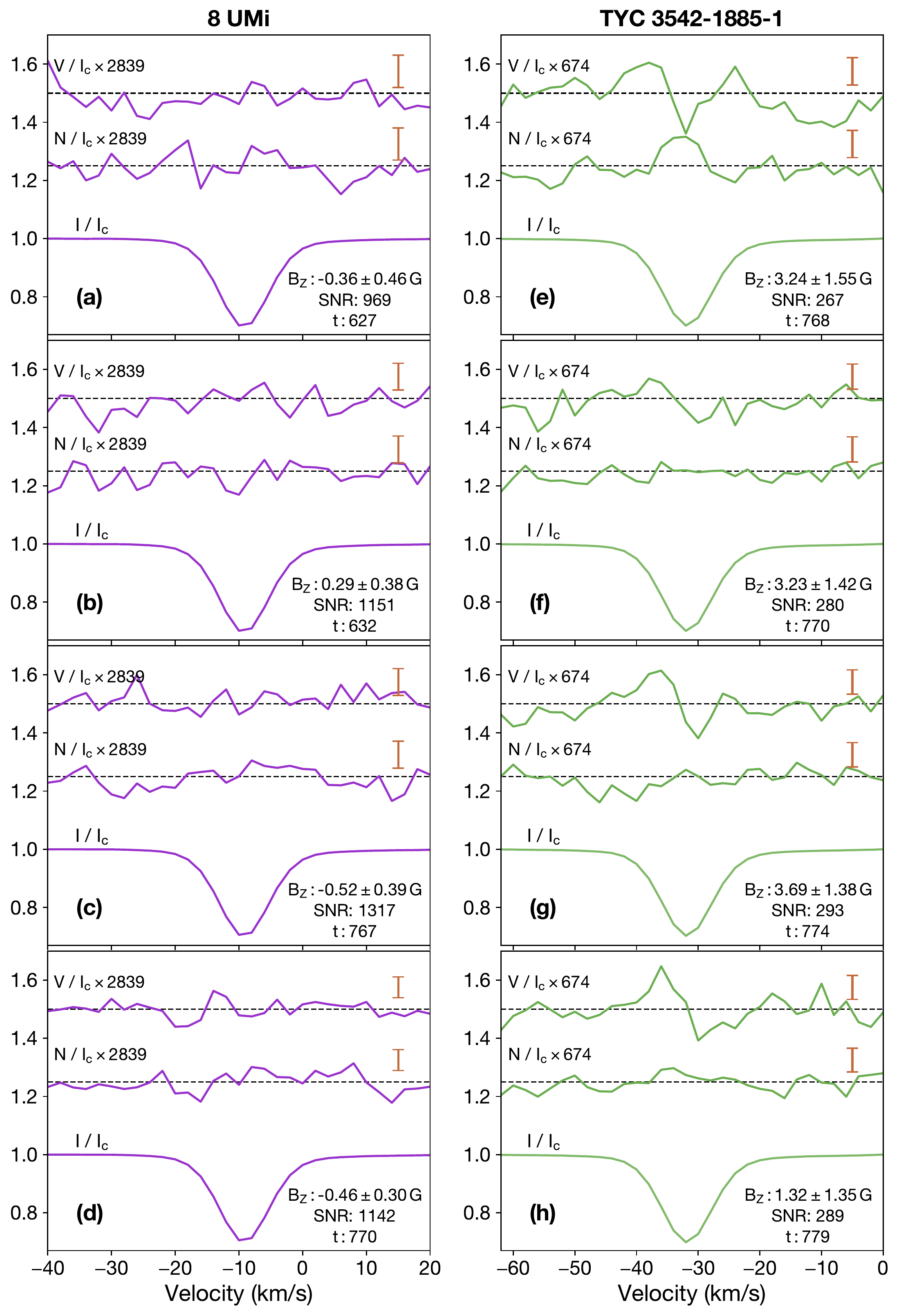}
\caption{\textbf{ESPaDOnS spectropolarimetry of the host star 8 UMi and active red giant TYC 3542-1885-1}
The least-squares deconvolution profiles in each panel, from top to bottom, are that of Stokes V, null polarisation $N$, and Stokes I, respectively.  Error bars indicate 1$\sigma$ (standard deviation) uncertainties for the profiles. 
Included are the Stokes $V$ mean longitudinal magnetic field strength ($B_Z$) and its corresponding 1$\sigma$ (standard deviation) uncertainty, polarimetric signal-to-noise ratio (SNR), and observation times ($t$) in BJD - 2459000.}
\label{fig:lsd_profiles}
\end{figure}

\begin{figure}[ht]
\centering
\includegraphics[width=1\linewidth]{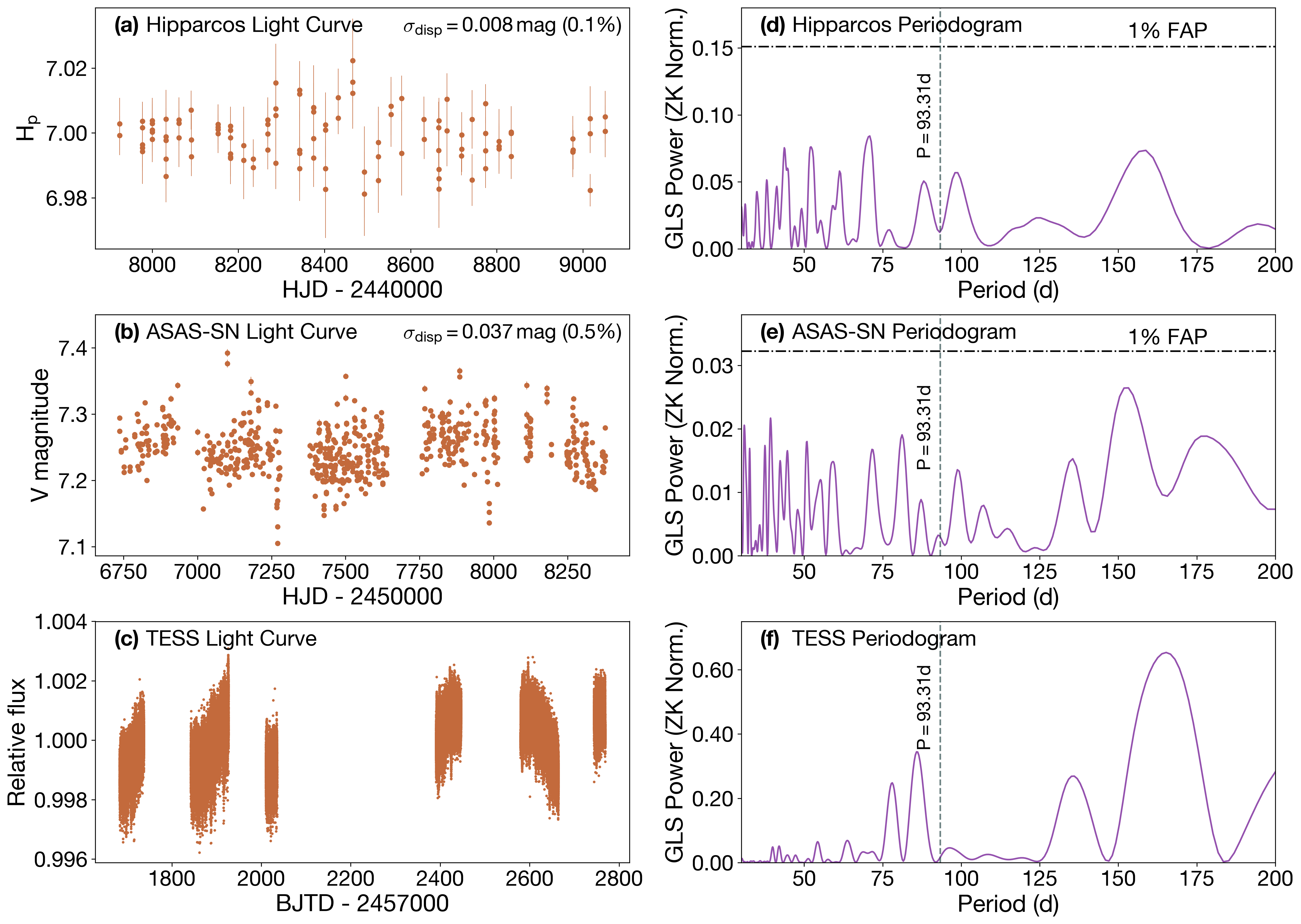}
\caption{ \textbf{Observed photometric variations of 8 UMi}.  Time series photometry 
from (\textit{a}) \textit{Hipparcos}, (\textit{b}) ASAS-SN, and (\textit{c}) systematics-corrected TESS Simple Aperture Photometry. The standard deviation uncertainty for each photometric measurement is shown with error bars. These are visible for the \textit{Hipparcos} data, but smaller than the symbol sizes for ASAS-SN and TESS data.
The dispersion of the \textit{Hipparcos} and ASAS-SN time series, $\sigma_{\mathrm{disp}}$, are quantified as a fraction of the star's apparent magnitude. Generalized Lomb Scargle (GLS) periodograms for the (\textit{d}) \textit{Hipparcos}, (\textit{e}) ASAS-SN, and (\textit{f}) systematics-corrected TESS Simple Aperture Photometry light curves. The vertical dashed line indicates 8 UMi b's orbital period, and the horizontal lines indicate the periodogram's False Alarm Probability (FAP). 
}
\label{fig:phot_var}
\end{figure}


\begin{figure}
    \centering
    \includegraphics[width=0.8\linewidth]{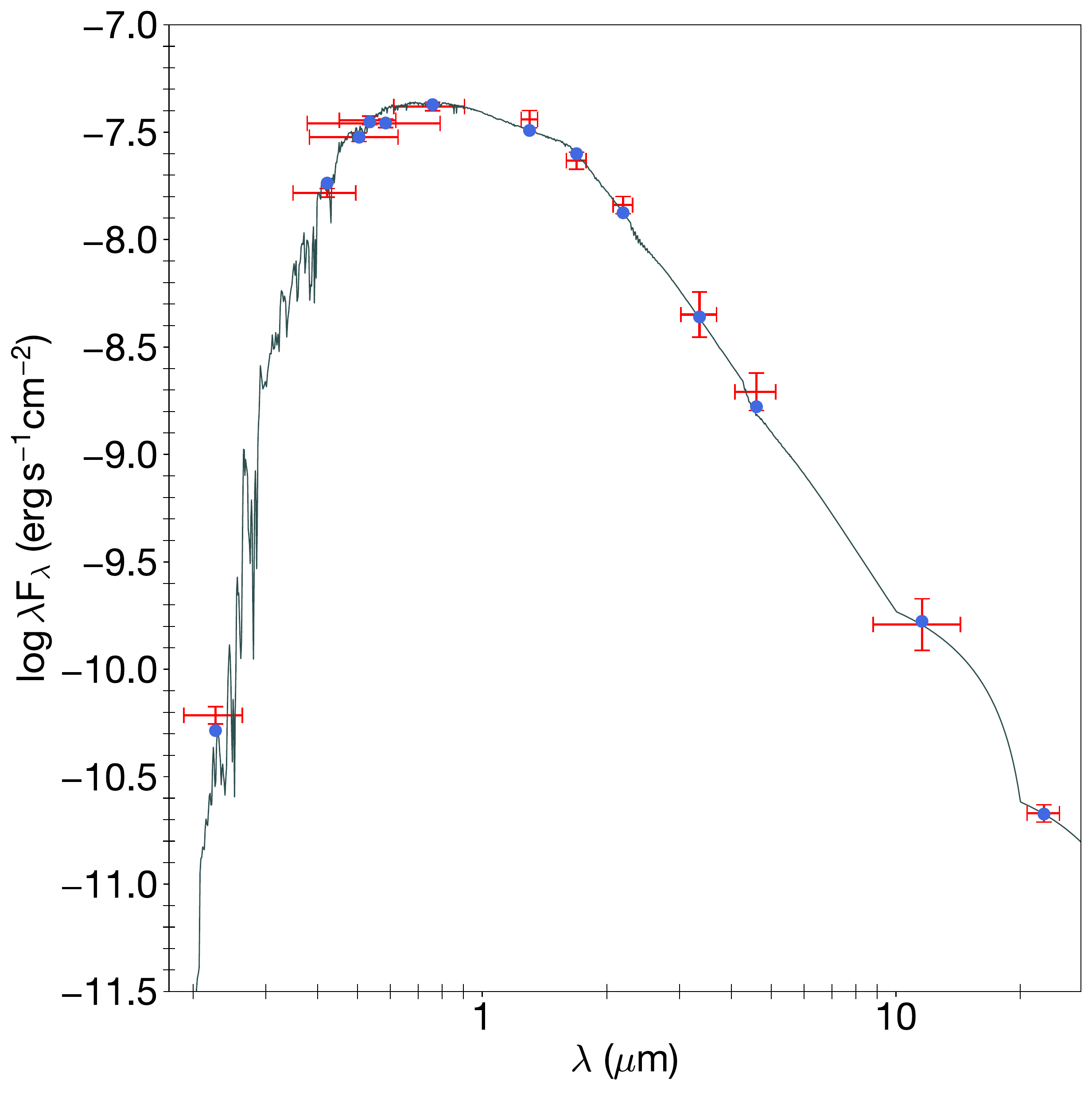}
    \caption{\textbf{Spectral energy distribution of 8~UMi}. The distribution was estimated using $B_T V_T$ magnitudes from {\it Tycho-2}, the $JHK_S$ magnitudes from {\it 2MASS}, the W1--W4 magnitudes from {\it WISE}, the $G G_{\rm BP} G_{\rm RP}$ magnitudes from {\it Gaia}, and the NUV magnitude from {\it GALEX}. Red symbols represent the observed photometric measurements, where the horizontal error bars represent the effective width of the passband while the vertical error bars are 1$\sigma$ (standard deviation) photometric uncertainties. Blue symbols are the model fluxes from the best-fit Kurucz atmosphere model (black), which have a reduced $\chi^2$ of 1.3, with extinction $A_V = 0.06 \pm 0.02$ mag, $T_{\rm eff} = 4900 \pm 75$~K, surface gravity $\log g = 2.5 \pm 0.5$ dex, and [Fe/H] = $-0.5 \pm 0.3$ dex. Integration of the (unreddened) model SED gives the bolometric flux at Earth, $F_{\rm bol} = 6.48 \pm 0.22 \times 10^{-8}$ erg~s$^{-1}$~cm$^{-2}$. \label{fig:sed}}
\end{figure}

\begin{figure}[ht]
\centering
\includegraphics[width=0.8\linewidth]{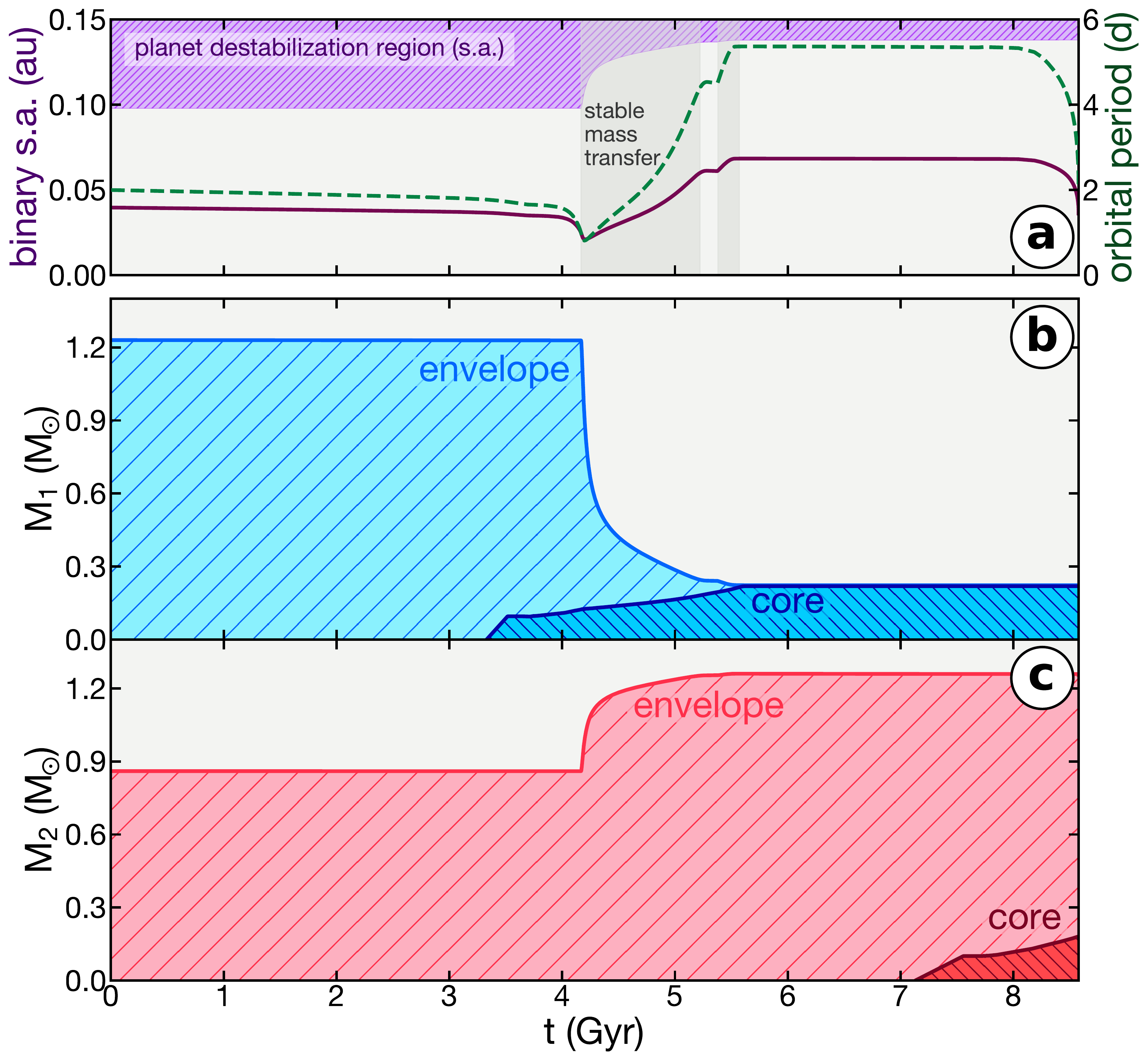}
\caption{ \textbf{Simulation of a stellar binary history for 8 UMi leading up to a stellar merger}. This fiducial model is simulated using $\beta=0.6$, $q=0.7$, and $P_{\mathrm{init}}=2\,\mathrm{d}$, with the stellar merger occuring at the onset of unstable mass transfer at $t\approx8.6\, {\rm Gyr}$. (\nzrremove{\textit{Top}}\nzradd{\textit{a}}) \nzrremove{Orbital s}\nzradd{Binary separation} (semi-major axis, \textit{s.a.}) \nzradd{(\textit{purple solid line}) and orbital period (\textit{green dashed line})} versus time for the simulated binary model\nzradd{.} (\nzrremove{\textit{Middle}}\nzradd{\textit{b}}) Primary total and core masses versus time. (\nzrremove{\textit{Bottom}}\nzradd{\textit{c}}) Secondary total/core mass versus time.}
\label{fig:time_plot}
\end{figure}

\begin{figure}[ht]
\centering
\includegraphics[width=0.8\linewidth]{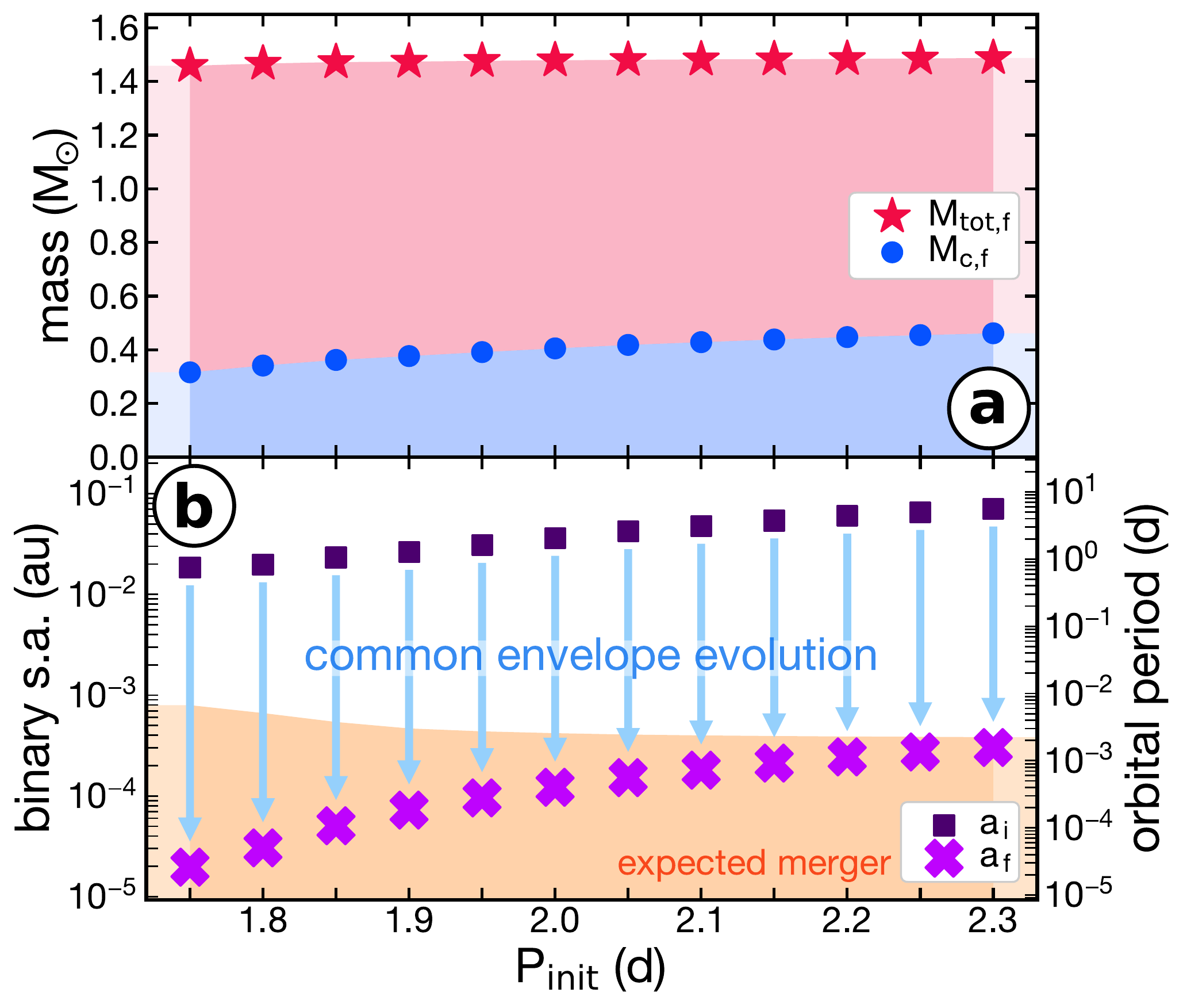}
\caption{ \textbf{Simulated white dwarf--red giant binaries that successfully merge to produce a core-helium burning giant like 8 UMi.} These models are simulated using $\beta=0.6$ and $q=0.7$, with $M_1=1.23\,M_\odot$ and $M_2=0.86\,M_\odot$. (\nzrremove{\textit{Top}}\nzradd{\textit{a}}) Final total mass $M_{\mathrm{tot,f}}$ and final \nzradd{helium} core mass $M_{\mathrm{c,f}}$ of the merger remnant.
(\nzrremove{\textit{Bottom}}\nzradd{\textit{b}}) Binary\nzrremove{orbital} separation (semi-major axis, \textit{s.a.}) \nzradd{and orbital period} before ($a_{\mathrm{i}}$) and after ($a_{\mathrm{f}}$) the common-envelope event.
The orange region represents $a_{\mathrm{f}}$ values that result in a stellar merger.}
\label{fig:magnetic_braking}
\end{figure}

\setcounter{table}{0}
\renewcommand{\tablename}{Extended Data Table}
\renewcommand{\thetable}{\arabic{table}}

\begin{table}[ht]

\caption{\label{tab:rv} Radial velocity measurements of 8 UMi collected using the HIRES spectrograph atop the Keck-I telescope from 19 May 2021 to 1 August 2022. Reported uncertainties are 1$\sigma$ (standard deviation) uncertainties.}
\begin{minipage}{0.495\textwidth}
\centering
\begin{tabular}{ccc}
\toprule
Time & Radial velocity & Uncertainty \\
BJD - 2459000 & m/s & m/s \\
\midrule
353.842103 &  34.810319 & 0.957495 \\
361.902204 &  69.348529 & 0.888507 \\
368.784719 &  48.809886 & 0.911906 \\
373.801558 &  51.160183 & 0.869466 \\
378.834098 &   5.876205 & 0.842353 \\
378.835070 &   7.768937 & 0.847276 \\
378.835996 &   4.595331 & 0.859393 \\
378.870140 &  -0.197723 & 0.894331 \\
378.871077 &  -1.580994 & 0.891074 \\
378.872026 &  -4.547195 & 0.912844 \\
378.979199 &  14.606174 & 0.857701 \\
378.980125 &  17.859138 & 0.896076 \\
378.981074 &  16.502770 & 0.848254 \\
384.890366 &  -2.409111 & 0.990540 \\
384.892484 &  -4.921133 & 0.958773 \\
384.894429 &  -2.751882 & 0.903895 \\
389.914801 & -19.275643 & 0.834250 \\
389.915738 & -17.315992 & 0.936139 \\
389.916674 & -15.740141 & 0.951262 \\
395.864380 & -39.513886 & 0.955047 \\
395.865318 & -37.492119 & 0.956622 \\
395.866209 & -39.848328 & 0.941511 \\
399.827438 & -31.820359 & 0.972085 \\
399.828607 & -34.245779 & 1.021377 \\
399.829741 & -33.843625 & 0.971329 \\
406.739231 & -60.907537 & 1.026887 \\
406.740134 & -60.100727 & 0.900744 \\
406.741048 & -58.346060 & 0.952770 \\
415.752084 & -27.386193 & 1.087321 \\
415.753681 & -30.211521 & 1.063041 \\
415.754700 & -26.923308 & 1.070932 \\
415.757998 & -27.737135 & 1.078141 \\
420.829269 & -49.413959 & 1.134881 \\
420.830415 & -44.291389 & 1.021217 \\
420.831688 & -45.008137 & 1.095143 \\
435.773916 &  19.435892 & 1.090663 \\
435.774865 &  19.687407 & 1.025351 \\
435.775814 &  18.302992 & 1.008119 \\
443.730761 &  24.779948 & 1.039300 \\
443.731675 &  25.534491 & 1.101412 \\
443.732636 &  23.593110 & 1.041690 \\
450.766041 &  69.011527 & 1.207604 \\
450.767025 &  68.430295 & 1.194408 \\
450.768009 &  69.318575 & 1.264927 \\
455.740426 &  62.917694 & 1.017182 \\
455.741352 &  56.633304 & 1.050469 \\
455.742266 &  58.825162 & 1.011381 \\
470.711559 &  37.405981 & 1.119984 \\
470.712497 &  35.467619 & 1.096883 \\
\bottomrule
\end{tabular}
\end{minipage}
\begin{minipage}{0.495\textwidth}
\centering
\begin{tabular}{ccc}
\toprule
Time & Radial velocity & Uncertainty \\
BJD - 2459000 & m/s & m/s \\
\midrule
470.713446 &  35.582958 & 1.120459 \\
478.709022 &  -4.443548 & 1.153627 \\
478.715029 &  -7.003524 & 1.268035 \\
478.720607 &  -7.905698 & 1.112496 \\
483.753253 & -50.993642 & 1.066348 \\
483.754468 & -52.206761 & 1.257434 \\
483.755672 & -49.098157 & 1.205983 \\
489.704354 & -67.520930 & 1.201372 \\
489.705523 & -70.438690 & 1.325950 \\
489.706692 & -71.966822 & 1.216982 \\
497.701247 & -59.438410 & 1.199675 \\
497.702809 & -59.489978 & 1.199892 \\
506.689152 & -57.215682 & 1.253242 \\
506.690240 & -59.439143 & 1.182603 \\
506.691351 & -59.210949 & 1.178573 \\
513.697277 & -46.329758 & 1.231641 \\
513.698481 & -48.825172 & 1.310779 \\
513.699719 & -49.372107 & 1.139539 \\
622.113710 &  28.831742 & 1.029991 \\
622.114740 &  26.803767 & 1.150842 \\
622.115794 &  29.397291 & 1.097965 \\
626.086114 &  29.669555 & 1.173846 \\
626.087202 &  29.863729 & 1.058872 \\
626.088325 &  30.465299 & 1.134471 \\
632.073919 &  42.695454 & 0.957066 \\
632.074868 &  44.096309 & 0.999416 \\
632.075828 &  43.932892 & 0.996506 \\
655.010662 &  43.071189 & 0.939295 \\
655.011599 &  41.502631 & 0.898865 \\
655.012525 &  43.082613 & 0.915619 \\
661.035709 &  33.081970 & 0.956038 \\
661.036901 &  37.172227 & 0.894799 \\
661.038059 &  32.319343 & 0.939563 \\
672.030706 &  -1.852284 & 0.967251 \\
672.031759 &  -3.129297 & 1.031975 \\
672.032870 &  -4.028580 & 0.936177 \\
681.007997 & -43.712090 & 0.982126 \\
681.008888 & -48.960588 & 0.960503 \\
681.009791 & -47.199496 & 1.010990 \\
691.000779 & -48.491071 & 1.150332 \\
691.001717 & -46.239763 & 1.064292 \\
691.002654 & -45.255098 & 0.997759 \\
695.013168 & -47.220558 & 1.019411 \\
695.014082 & -43.609635 & 1.043123 \\
695.014996 & -45.622370 & 0.948686 \\
700.948879 & -20.122784 & 1.002295 \\
700.952375 & -19.824926 & 1.230983 \\
711.897946 &  13.432848 & 0.980607 \\
711.898907 &  11.327911 & 0.985397 \\
\bottomrule
\end{tabular}
\end{minipage}
\end{table}

\setcounter{table}{0}
\renewcommand{\tablename}{Extended Data Table}
\renewcommand{\thetable}{\arabic{table} (continued)}

\begin{table}[ht]
\centering
\caption{ }

\begin{tabular}{ccc}
\toprule
Time & Radial velocity & Uncertainty \\
BJD - 2459000 & m/s & m/s \\
\midrule
711.899856 &  13.267244 & 1.085308 \\
715.974755 &  37.686341 & 0.961920 \\
715.975693 &  39.381356 & 0.998087 \\
715.976630 &  37.921361 & 0.917527 \\
737.813506 &  76.820012 & 0.909781 \\
737.814536 &  76.766067 & 0.996631 \\
737.815566 &  74.604152 & 0.918539 \\
740.799098 &  62.368957 & 0.937384 \\
740.800070 &  65.637292 & 0.875799 \\
740.801019 &  67.465725 & 0.915763 \\
741.796518 &  54.978361 & 0.888018 \\
741.797409 &  51.933870 & 0.965281 \\
741.798323 &  53.079849 & 0.865434 \\
747.917120 &  33.196888 & 0.951404 \\
747.918335 &  33.229634 & 0.994584 \\
747.919574 &  36.687595 & 0.950376 \\
749.986941 &  23.051076 & 0.871991 \\
749.988017 &  18.465904 & 0.846402 \\
749.989071 &  23.438575 & 0.963685 \\
756.960219 &   2.667942 & 0.956022 \\
756.961353 &  -1.127702 & 0.926351 \\
756.962418 &  -0.344195 & 0.903171 \\
759.944743 & -14.055632 & 0.993686 \\
759.945773 & -14.769997 & 0.991786 \\
759.946757 & -15.594268 & 0.884224 \\
771.854751 & -50.035824 & 0.973484 \\
771.855862 & -48.483816 & 1.025444 \\
771.856973 & -49.680070 & 1.070451 \\
789.800968 & -11.766739 & 1.007803 \\
789.802218 & -10.835423 & 1.091427 \\
789.803537 &  -7.329680 & 1.093144 \\
791.806519 & -13.067793 & 1.080876 \\
791.807607 & -10.685886 & 1.084655 \\
791.808834 & -10.084930 & 1.104502 \\
792.739494 & -29.757211 & 1.138764 \\
792.740535 & -28.673963 & 1.135497 \\
792.741589 & -27.545748 & 1.123028 \\
\bottomrule
\end{tabular}
\end{table}

\setcounter{table}{1}
\renewcommand{\tablename}{Extended Data Table}
\renewcommand{\thetable}{\arabic{table}}

\clearpage
\bibliographystyleSupp{naturemag}
\bibliographySupp{sample}

\end{document}